\documentclass[aps,prb,twocolumn,10pt,a4paper]{revtex4-1}
\usepackage{graphicx}
\usepackage{amsmath}
\usepackage{amssymb}
\usepackage{bm}

\begin{document}


\newcommand{\tk}{T_\mathrm{K}}
\newcommand{\boldk}{{\boldsymbol{\mathrm{k}}}}
\newcommand{\K}{\boldk}
\newcommand{\modmutil}{|\tilde{\mu}|}
\newcommand{\modmu}{|\mu|}
\newcommand{\hfmu}{|\mu_0|}
\newcommand{\bra}[1]{\langle #1 |}
\newcommand{\ket}[1]{| #1 \rangle}
\newcommand{\cre}[1]{c_{#1}^\dagger}
\newcommand{\ann}[1]{c_{#1}^{\phantom{\dagger}}}
\newcommand{\dbyd}[2]{\left(\frac{\partial #1}{\partial #2}\right)}
\newcommand{\sgn}{\operatorname{sgn}}
\newcommand{\E}{\mathrm{e}}
\newcommand{\D}{\;\mathrm{d}}
\newcommand{\I}{\mathrm{i}}
\newcommand{\ex}[1]{\langle #1 \rangle}
\newcommand{\tr}{\operatorname{Tr}}
\newcommand{\brapn}{\bra{\Psi_0^N}}
\newcommand{\ketpn}{\ket{\Psi_0^N}}
\newcommand{\deltar}{\Delta_\mathrm{R}}
\newcommand{\deltai}{\Delta_\mathrm{I}}
\newcommand{\deltao}{\Delta_0}
\newcommand{\sigr}{\Sigma^\mathrm{R}}
\newcommand{\sigi}{\Sigma^\mathrm{I}}
\newcommand{\gupa}{G_\uparrow^\mathrm{A}(\omega)}
\newcommand{\gupb}{G_\uparrow^\mathrm{B}(\omega)}
\newcommand{\gdna}{G_\downarrow^\mathrm{A}(\omega)}
\newcommand{\gdnb}{G_\downarrow^\mathrm{B}(\omega)}
\newcommand{\ga}[1]{G_{#1}^\mathrm{A}(\omega)}
\newcommand{\gb}[1]{G_{#1}^\mathrm{B}(\omega)}
\newcommand{\half}{\frac{1}{2}}
\newcommand{\niup}{\hat n_{\mathrm{i}\uparrow}}
\newcommand{\nidn}{\hat n_{\mathrm{i}\downarrow}}
\newcommand{\nisig}{\hat n_{\mathrm{i}\sigma}}
\newcommand{\scrgu}{\mathcal{G}_\uparrow}
\newcommand{\scrgd}{\mathcal{G}_\downarrow}
\newcommand{\ut}{\tilde U}
\newcommand{\dc}{\tilde \delta_\mathrm{c}}
\newcommand{\dt}{\tilde \delta}
\newcommand{\ot}{\tilde \omega}
\newcommand{\omegam}{\omega_\mathrm{m}}
\newcommand{\hc}{\hat{H}_\mathrm{c}}
\newcommand{\hh}{\hat{H}_\mathrm{h}}
\newcommand{\hp}{\hat{H}_\mathrm{p}}
\newcommand{\mub}{\mu_\mathrm{B}}
\newcommand{\epsi}{\epsilon_\I}
\newcommand{\epsit}{\tilde\epsi}
\newcommand{\simutoinf}{\overset{U\to\infty}{\sim}}
\newcommand{\simuttoinf}{\overset{\tilde U\to\infty}{\sim}}
\newcommand{\omo}{\ot^0_\mathrm{m}}
\newcommand{\om}{\omega_\mathrm{m}}
\newcommand{\gt}{\tilde\Gamma}
\newcommand{\ua}{\uparrow}
\newcommand{\da}{\downarrow}
\newcommand{\et}{\tilde \epsilon}
\newcommand{\thalf}{\tfrac{1}{2}}


\newcommand{\comment}[1]{}

\newcommand{\w}{\omega}
\newcommand{\wpr}{\omega^{\prime}}

\newcommand{\Ss}{\Sigma_{\sigma}}
\newcommand{\Ssp}{\Sigma_{\sigma}^{\prime}}
\newcommand{\SsR}{\Sigma_{\sigma}^{R}}
\newcommand{\SsI}{\Sigma_{\sigma}^{I}}
\newcommand{\Gs}{G_{\sigma}}

\newcommand{\tI}{\tilde{I}_{\sigma\sigma^{\prime}}}
\newcommand{\tgamma}{\tilde{\Gamma}_{\sigma\sigma^{\prime}}}
\newcommand{\pd}{\phantom\dagger}
\newcommand{\chic}{\chi^{\pd}_{\mathrm{c,imp}}}
\newcommand{\chis}{\chi^{\pd}_{\mathrm{s,imp}}}
\newcommand{\tm}{\tilde{\mu}}
\newcommand{\chisloc}{\chi^{\pd}_{s}}
\newcommand{\nimp}{n_{\mathrm{imp}}}
\newcommand{\nimps}{n_{\mathrm{imp},\sigma}}
\newcommand{\nimpas}{n_{\mathrm{imp},A\sigma}}
\newcommand{\mimp}{m_{\mathrm{imp}}}
\newcommand{\mimpa}{m_{\mathrm{imp},A}}
\newcommand{\rhoimps}{\Delta\rho_{\mathrm{imp},\sigma}(\w, h)}


\title{Mott insulators and the doping-induced Mott transition within DMFT: exact results for the one-band Hubbard model.}


\author{David E. Logan and Martin R. Galpin}
\affiliation{Department of Chemistry, Physical and Theoretical Chemistry, Oxford University, South Parks Road, Oxford, OX1 3QZ, United Kingdom}


\begin{abstract}
The paramagnetic phase of the one-band Hubbard model is studied at zero-temperature, within the framework of dynamical mean-field theory, and for general particle-hole asymmetry where  a doping-induced Mott transition occurs. Our primary focus is the Mott insulator (MI) phase, and our main aim to establish what can be shown exactly about it. To handle the locally doubly-degenerate MI requires two distinct self-energies, which reflect the broken symmetry nature of the phase and together determine the standard single self-energy. Exact results are obtained for the local charge, local magnetic moment and associated spin susceptibilities, the interaction-renormalised levels, and the low-energy behaviour of the self-energy in the MI phase. The metallic phase is also considered briefly, and shown to acquire an emergent particle-hole symmetry as the Mott transition is approached. Throughout the metal, Luttinger's theorem is reflected in the vanishing of the Luttinger integral;
for the generic MI by contrast this is shown to be non-vanishing, but again to have a universal magnitude. 
Numerical results are also obtained using NRG, for the metal/MI phase boundary, the scaling behaviour of the charge
as the Mott transition is aproached from the metal, and associated universal scaling of single-particle dynamics as
the low-energy Kondo scale vanishes.
\end{abstract}

\maketitle

\section{Introduction}
\label{section:intro}

Understanding the interaction-driven Mott insulating state remains a central challenge in condensed matter science; playing a major role in a host of materials, including transition-metal oxide and related compounds, where the Mott transition to it from a metallic phase can be induced e.g.\ by chemical doping or applied pressure.

The simplest model to capture a Mott insulator and the attendant Mott transition, is of course the 
Hubbard model:~\cite{Hubbard1963,*Gutzwiller1963,*Kanamori1963} a one-band tight-binding model supplemented 
by a local Coulomb repulsion $U$ which drives the transition. Considerable progress in understanding 
the model -- and correlated lattice-fermions in general -- has arisen with the advent of dynamical mean-field 
theory~\cite{MetznerVollhardt1989,MuellerH1989a,*MuellerH1989b,GeorgesKotliarPRB1992,JarrellPRL1992}
(DMFT, for a review see ref.\ \onlinecite{dmftgeorgeskotliar}). Formally exact in the limit of infinite dimensionality 
or coordination number, and characterised as such by a purely local (momentum-independent) interaction self-energy, DMFT
is known to capture well many properties of real materials.~\cite{KotVollPhyToday2004,*KotliarEtAlRMP2006}

One highlight of this approach has been a detailed understanding of the Mott transition in the Hubbard model 
without magnetic ordering (either by neglecting it, or equivalently by ensuring its absence through frustration). 
While early studies focused mainly on the particle-hole (ph) symmetric limit,~\cite{ dmftgeorgeskotliar} 
where the local charge $n=1$ for all $U$ throughout both phases, a large body of work has been devoted to 
the problem away from ph-symmetry;\cite{dmftgeorgeskotliar,PruschkeCoxJarrellEPL1993,*JarrellPruschkePRB1994,*PruschkeJarrellFreericksAdvPhys1995,FisherKotliar1995,*KajueterKotliarPRL1996,*KajueterKotliar1996,MerinoMcKenzie2000,KotMurthyRozenPRL2002,WernerMillis2007,HallbergRozenberg2007,Zitko+BSS2013,ZitKotGeorgesRQP2013,RozenbergDobrosPRL2015} 
where the resultant doping-induced Mott transition arises as the carrier concentration $\delta =|1-n|$ vanishes, with $n=1$ throughout the Mott insulator. The problem is highly rich, and continues to yield new insights; recent 
discoveries include the resilient persistence of quasiparticles to temperatures ($T$) well above those characteristic of 
low-energy Fermi liquid behaviour,\cite{ZitKotGeorgesRQP2013} and an intimate connection between Mott quantum criticality and the `bad metal' behaviour reflected in a linear $T$-dependence of resistivity.~\cite{RozenbergDobrosPRL2015}

Since any lattice-fermion model within DMFT reduces to an effective, local quantum impurity model coupled to a 
self-consistently determined `host',~\cite{dmftgeorgeskotliar} Kondo physics in one form or another is involved.
For the metallic phase of the Hubbard model, the standard Kondo effect occurs, quenching fully the electron spin degrees of freedom and producing a non-degenerate ground state. For the Mott insulator by contrast, the `host' spectrum is gapped and the ground state characterised by an entropy of $k_{B}\ln2$ per site; which local double-degeneracy reflects incomplete 
spin-quenching, and hence a residual local moment.

Most work on the Hubbard model has tended to focus largely on the metallic phase, and the approach to the Mott transition from it, associated with a vanishing low-energy Kondo scale $\w_{\mathrm{K}}$ and collapse of the Kondo resonance in the local single-particle spectrum.~\cite{dmftgeorgeskotliar} The metal is of course perturbatively connected to the non-interacting limit of the model, and as such is a Fermi liquid, in which Luttinger's theorem holds.
 This in turn enables a number of exact results to be obtained, from the low-energy theory of the metallic Anderson impurity model onto which the problem maps.~\cite{dmftgeorgeskotliar}

The Mott insulator by contrast is not adiabatically connected to the non-interacting limit, is not in consequence
a Fermi liquid, and the usual Luttinger theorem does not hold. An obvious question then is: what can be deduced exactly about the Mott insulating phase? This is our primary focus here, considering $T=0$ where the distinction between metallic and insulating phases is sharp.

Answers to the question are obtained by exploiting the DMFT mapping onto an effective impurity model, coupled
with recent work on non-Fermi liquid phases in quantum impurity models.~\cite{LTG2014} To handle the locally doubly degenerate Mott insulator requires two distinct self-energies, which reflect the broken symmetry character of that phase and are themselves  directly calculable from many-body perturbation theory,~\cite{LTG2014} as functional derivatives of a 
Luttinger-Ward functional. These two self-energies together determine the conventional single
self-energy -- usually thought of as `the' self-energy. We remark  here that, while a two-self-energy (TSE) description also underlies the local moment approach\cite{LET,*MTGLMA_asym,*nigelscalspec,fnrefLMAPAM} to correlated electrons, 
its use here is exact~\cite{LTG2014} and unencumbered by any subsequent approximations.

Following a summary of the model and DMFT background (sec.\ \ref{section:sec2}), 
we first consider briefly two exact results for the metallic phase (sec.\ \ref{section:sec3}); which
together reveal the existence of an \emph{emergent} ph-symmetry as the doping induced Mott transition is approached from the metal, for the generic ph-asymmetric model.

We turn specifically to the Mott insulator in sec.\ \ref{section:sec4}, beginning with an overview of essential elements of the TSE description.~\cite{LTG2014} Results are then obtained for the local charge (sec.\ \ref{subsection:sec4A}), and the local magnetic moment and associated spin susceptibilites (sec.\ \ref{subsection:sec4B});
expressed in terms of spin-dependent renormalised levels -- effective single-particle levels renormalised by electron interactions, of a type familiar in quantum impurity physics,~\cite{hewsonbook} but here associated with the \emph{two} 
self-energies. Their properties are determined in sec.\ \ref{subsection:sec4C}.
The behaviour of the conventional single self-energy is considered in sec.\ \ref{subsection:sec4D}; 
showing in particular that, at least sufficiently close to ph-symmetry, the self-energy contains a low-energy pole inside the insulating gap away from the Fermi energy, which persists robustly down to the Mott transition.

In sec.\ \ref{subsection:sec4E} we consider the standard Luttinger integral expressed in terms of the single self-energy, the vanishing of which for all interactions $U$ throughout the metallic phase is tantamount to Luttinger's theorem. This result does not hold in the Mott insulator. It is nevertheless shown that, for the generic ph-asymmetric model, the magnitude of the Luttinger integral again has a constant value -- now non-vanishing -- for all $U$ throughout the Mott insulator; and as such is an intrinsic hallmark of this phase, in the same sense that its vanishing throughout the metal is characteristic of the Fermi liquid.

The theory developed is shown as we go to be fully supported by numerics, using numerical renormalisation group (NRG) calculations. Sec.\ \ref{section:sec5} is devoted to further results obtained via NRG. Throughout the paper the model's 
ph-asymmetry is parameterised by $\eta = 1+2\epsilon_{d}/U$ ($\epsilon_{d}$ is the site-energy), with $\eta =0$ the 
ph-symmetric limit; this is partly for convenience, since as shown in sec.\ \ref{subsection:PD} the Mott insulating phase arises only for $|\eta|<1$. NRG results are obtained for the metal/Mott insulator phase boundary in the $(U,\eta)$-plane, 
including its functional form; together with both the critical behaviour and scaling form~\cite{WernerMillis2007} of the charge (and charge susceptibility) as the Mott transition is approached from the metallic side.
Finally, in sec.\ \ref{subsection:sec5A} we consider the single-particle spectrum $D(\w)$ as the transition is approached 
from the metal and the low-energy Kondo scale $\w_{\mathrm{K}}$ vanishes. Clear universal scaling of $D(\w)$  as 
a function of $\w/\w_{\mathrm{K}}$ is found, \emph{provided} the critical $U$ for the transition is approached at any fixed asymmetry $|\eta| <1$; and which explains the absence of universality, except on the lowest energy scales 
$|\w|/\w_{\mathrm{K}} \ll 1$, recently found in ref.\ \onlinecite{Zitko+BSS2013}.


\section{Model and background}
\label{section:sec2}

We consider the one-band Hubbard model 
\begin{equation}
\label{eq:1}
\hat{H} ~=~ \sum_{i} \Big(\epsilon_{d}^{\pd}\hat{n}_{i}^{\pd} + U\hat{n}_{i\uparrow}^{\pd}
\hat{n}_{i \downarrow}^{\pd}\Big)~-~t\sum_{(i,j),\sigma} c_{i\sigma}^{\dagger}c_{j\sigma}^{\pd},
\end{equation}
with $\hat{n}_{i\sigma} =c_{i\sigma}^{\dagger}c_{i \sigma}^{\pd}$, $\hat{n}_{i} =\sum_{\sigma}\hat{n}_{i\sigma}$ the local  
number operator, and the $(i,j)$ sum over nearest neighbour (NN) lattice sites.
$U$ is the on-site Coulomb interaction, and $\epsilon_{d}$ the one-electron site-energy (alternatively, with 
$\epsilon_{d}\equiv -\mu$, one may work instead with a chemical potential $\mu$).
The NN hopping is rescaled within DMFT as~\cite{dmftgeorgeskotliar} $t= t_{*}/(2\sqrt{Z_{c}})$ with coordination number 
$Z_{c} \rightarrow \infty$; and the only relevant property of the non-interacting energy dispersion is $\rho_{0}(\epsilon)$, the free density of states for $\epsilon_{d}=0$. This we take to be of standard bounded, 
semicircular form
\begin{equation}
\label{eq:2}
\rho_{0}^{\pd}(\epsilon)~=~ \frac{2}{\pi t_{*}^{2}} \sqrt{ t_{*}^{2}- \epsilon^{2}}
\end{equation}
with band halfwidth $t_{*}$ (corresponding formally to a Bethe lattice).

 The model can thus be parameterised by $\epsilon_{d}$ and $U$. Equivalently, it can also be specified by $U$ and the asymmetry $\eta$, defined by
\begin{equation}
\label{eq:3}
\eta~=~ 1~+~\frac{2\epsilon_{d}^{\pd}}{U}
\end{equation}
with $\eta =0$ at the  ph-symmetric point of the model, $\epsilon_{d} =-U/2$. 
Under a ph-transformation it is easily shown that (a) $\hat{H}(\eta,U) \equiv \hat{H}(-\eta,U)$, so that only 
e.g.\ $\eta \geq 0$ need be considered; and (b) the mean charge per site, 
$n=\sum_{\sigma}\langle \hat{n}_{i\sigma}\rangle$, satisfies $n(\eta,U) - 1 = -[n(-\eta,U)-1]$. Hence $n=1$ for all $U$ 
at ph-symmetry; while $n\leq 1$ for all $\eta \geq 0$ and any $U$.
The charge is of course related to the local
retarded propagator $G(\w)$ 
($ \leftrightarrow G(t) = -i\theta(t)\langle \{c_{i\sigma}^{\pd}(t), c_{i\sigma}^{\dagger}\}\rangle$) by
\begin{equation}
\label{eq:4}
\tfrac{1}{2}n ~=~ -\tfrac{1}{\pi}\mathrm{Im} \int^{0}_{-\infty}d\w ~G(\w),
\end{equation}
with $\w =0$ the Fermi level. \\

Within DMFT the propagator is given by~\cite{dmftgeorgeskotliar}
\begin{subequations}
\label{eq:5}
\begin{align}
G(\w)~=&~ \int^{\infty}_{-\infty}d\epsilon ~ \rho_{0}^{\pd}(\epsilon) G(\epsilon ; \w)
\\
=&~ \int^{\infty}_{-\infty}d\epsilon ~\frac{\rho_{0}^{\pd}(\epsilon)}{\w^{+} - \epsilon_{d}^{\pd} -\Sigma(\w) - \epsilon},
\end{align}
\end{subequations}
with $\Sigma(\w)=\Sigma^{R}(\w)-i\Sigma^{I}(\w)$ the interaction self-energy (purely local, independent of $\epsilon$),
and $\w^{+} = \w +i0+$. With $\rho_{0}(\epsilon)$ from eq.\ \ref{eq:2}, this may be written equivalently as
\begin{equation}
\label{eq:6}
G(\w)~=~ \left[\w^{+} -\epsilon_{d}^{\pd} -\Sigma(\w) - \tfrac{1}{4}t_{*}^{2}G(\w)
\right]^{-1}
\end{equation}
with local Feenberg self-energy $S(\w)=\tfrac{1}{4}t_{*}^{2}G(\w)$ 
(i.e.\ $S(\w)=\sum_{j} t^{2}G_{jj;\sigma}(\w)$ with sites $j$ NN to $i$). Eq.\ \ref{eq:6}
points up the fact that within DMFT any lattice-fermion model reduces  to a self-consistent quantum impurity 
problem;~\cite{dmftgeorgeskotliar} for it is precisely that for an Anderson impurity model coupled to a bath specified by a hybridization function `$\Gamma(\w)$' ($= \tfrac{1}{4}t_{*}^{2}G(\w)$) that must be self-consistently determined.
For self-consistent metallic solutions the effective impurity model is the standard metallic Anderson model; while for insulating solutions it is that of a gapped impurity model, since the spectral density of the hybridization is gapped around the Fermi level.


\subsection{Overview: phase diagram}
\label{subsection:PD}

Our focus is the $T=0$ paramagnetic phase of the model, with particular emphasis on the Mott insulating phase and the Mott transition (MT). Whether at ph-symmetry or away from it, the Mott insulator (MI) is of course characterised by a mean charge per site of $n =1$, and as such is inexorably half-filled.

\begin{figure}
\includegraphics{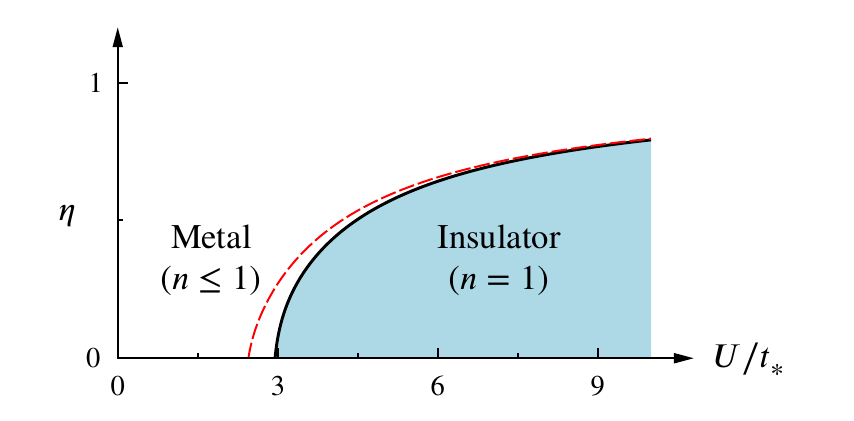}
\caption{\label{fig:fig1} 
Schematic phase boundary in the $(U,\eta)$-plane. Metal/MI phases are separated by
the solid line $\eta_{c}(U)$ (or equivalently $U_{c}(\eta)=U_{c}(-\eta)$); with charge $n=1$ throughout the MI, and $n\leq 1$ in the metal. Dashed line indicates $U_{c1}(\eta)$ (or $\eta_{c1}(U)$), the lower boundary to stability of insulating solutions. See text for discussion.
}
\end{figure}

 The phase diagram in the $(U,\eta)$-plane is shown schematically in Fig.\ \ref{fig:fig1}
(NRG results for it will be given in sec.\ \ref{section:sec5}). The metal/MI phase boundary is indicated by the solid 
line $\eta_{c}(U)$, or equivalently $U_{c}(\eta)$; and since $U_{c}(\eta) = U_{c}(-\eta)$, only $\eta \geq 0$ need be considered. Note that the MI occurs only for $|\eta| <1$ as explained below.

We first summarise some key qualitative features of the phases, and their single-particle dynamics, which are well known from previous studies 
(e.g.\ refs.\ \onlinecite{dmftgeorgeskotliar,PruschkeCoxJarrellEPL1993,*JarrellPruschkePRB1994,*PruschkeJarrellFreericksAdvPhys1995,FisherKotliar1995,*KajueterKotliarPRL1996,*KajueterKotliar1996,MerinoMcKenzie2000,KotMurthyRozenPRL2002,WernerMillis2007,HallbergRozenberg2007,Zitko+BSS2013,ZitKotGeorgesRQP2013,RozenbergDobrosPRL2015}):\\
\noindent (i) For given asymmetry $\eta <1$, self-consistent metallic solutions occur for all $U$ \emph{up} to 
$U_{c}(\eta)$ (a `$U_{c2}$' in traditional terminology). Insulating solutions by contrast persist 
for all $U$  \emph{down} to $U=U_{c1}(\eta)<U_{c}(\eta)$ (the dashed line in Fig.\ \ref{fig:fig1} shows $U_{c1}(\eta)$, or equivalently $\eta_{c1}(U)$). While metallic and insulating solutions thus coexist in the interval
$U_{c1}(\eta) <U<U_{c}(\eta)$, the metallic solutions have a lower energy. The metal-insulator transition thus occurs at 
$U=U_{c}(\eta)$ (Fig.\ \ref{fig:fig1}, solid line).\\
\noindent (ii) In the metallic phase sufficiently close to $U_{c}(\eta)$, the single-particle spectrum
$D(\w) =-\tfrac{1}{\pi}\mathrm{Im}G(\w)$ contains a Kondo resonance pinned to the Fermi level $\w =0$ (symptomatic of the standard metallic Kondo effect). The width of the Kondo resonance is characterised by a low-energy Kondo scale 
$\w_{\mathrm{K}}$  (proportional to the quasiparticle weight $Z = [1-(\partial\Sigma^{R}(\w)/\partial\w)_{\w =0}]^{-1}$).
This decreases progressively with increasing $U$ and vanishes continuously as $U \rightarrow U_{c}(\eta)-$ where the Kondo resonance vanishes `on the spot', leaving thereby the Mott insulating solution with a fully formed, finite spectral gap between the Hubbard bands in $D(\w)$. In otherwords, the Kondo resonance in the metal resides within a `preformed' insulating gap in $D(\w)$, which becomes the fully fledged insulating gap at $U=U_{c}(\eta)$. At ph-symmetry $\eta =0$, the 
resonance lies precisely in the middle of the preformed gap (as well known since early studies of the 
problem~\cite{dmftgeorgeskotliar,ZhangRosenKot1993,*RosenKotZhang1994}); while on increasing $\eta$ by contrast, 
it lies progressively closer to the upper edge of the lower Hubbard band in $D(\w)$.\\

As above, for given $\eta <1$ an insulating solution itself persists down to $U=U_{c1}(\eta) <U_{c}(\eta)$. An 
estimate of $U_{c1}(\eta)$ is obtained as follows. $D(\w)$ for the insulator consists of lower and upper Hubbard bands centred on $\w =\epsilon_{d}$ and $\epsilon_{d}+U$ respectively. Model the lower Hubbard band as a semicircular band centred on  
$\w =\epsilon_{d}$, the upper edge of which thus occurs at $\w_{+} = \epsilon_{d}+t_{*}$. If $\w_{+}$ lies below the Fermi level $\w =0$, the insulating solution is stable. The boundary to insulating stability  is thus $\epsilon_{d}=-t_{*}$; with a corresponding $\eta$ (eq.\ \ref{eq:3}) of $\eta_{c1}(U) = 1-2t_{*}/U \equiv 1-W/U$ (with $W=2t_{*}$ the full bandwidth), or equivalently $U_{c1}(\eta)/W = (1-\eta)^{-1}$. This simple result is in general only rough -- e.g.\ at ph-symmetry it gives
$U_{c1}(\eta =0)/W =1$, while NRG calculations yield $U_{c1}(0)/W \simeq 1.25$.~\cite{BullaPRL99}
It is however asymptotically exact as $U/t_{*} \rightarrow \infty$; for here double occupancy is strictly precluded, the upper Hubbard band is `projected out' to $\w \rightarrow \infty$, and the dynamics of the lower Hubbard band is that of a single hole in a random paramagnetic configuration of spins -- which within DMFT corresponds precisely to the non-interacting 
(semicircular) spectrum. Hence as $U/t_{*} \rightarrow \infty$,
\begin{equation}
\label{eq:7}
\eta_{c1}^{\pd}(U)~=~  1-\frac{2t_{*}}{U} ~\equiv ~1-\frac{W}{U}~~~~~~:~\frac{U}{t_{*}}\rightarrow \infty ~.
\end{equation}
As above moreover, $U_{c1}(\eta) < U_{c}(\eta)$, or equivalently $\eta_{c}(U) <\eta_{c1}(U)$ for any finite $U >U_{c}(0)$.
Hence from eq.\ \ref{eq:7}, $\eta_{c}(U) <1$ for any finite $U$, and the MI phase is indeed bounded by $\eta =1$.
Further, given that~\cite{FisherKotliar1995,*KajueterKotliar1996} the metallic Kondo resonance in $D(\w)$ as 
$U\rightarrow U_{c}(\eta)$ lies at most a finite distance ${\cal{O}}(t_{*})$ above the upper edge of the lower Hubbard band for $U/t_{*} \gg 1$, it follows that $\eta_{c}(U)$ is likewise of form $\eta_{c}(U) = 1-bW/U$ (with $b \geq 1$) as 
$U/t_{*}\rightarrow \infty$. Both $\eta_{c}(U)$ and $\eta_{c1}(U)$ thus tend asymptotically to unity in this limit.


\section{Metallic Fermi liquid}
\label{section:sec3}

Before turning to the MI, we consider briefly two exact results (eqs.\ \ref{eq:8},\ref{eq:12} below) for the metallic phase; 
which reflect its Fermi liquid character and, together, indicate the existence of an emergent particle-hole symmetry as the transition is approached from the metallic phase.

Since the metallic phase is a Fermi liquid, the imaginary part of the self-energy 
$\Sigma(\w)=\Sigma^{R}(\w)-i\Sigma^{I}(\w)$ vanishes at the Fermi level, $\Sigma^{I}(\w =0)=0$. From
eq.\ \ref{eq:5}b it follows trivially that the single-particle spectrum at the Fermi level is given by
\begin{equation}
\label{eq:8}
D(0)~=~\rho_{0}(-\epsilon_{d}^{*})
\end{equation}
where
\begin{equation}
\label{eq:9}
\epsilon_{d}^{*}~=~ \epsilon_{d}^{\pd}~+~\Sigma^{R}(0)
\end{equation}
is the interaction-renormalised level energy. 

 We return to eq.\ \ref{eq:8} below, but first obtain a result for the charge $n$. Using the obvious identity
$[\w^{+} - \epsilon_{d} - \Sigma(\w) -\epsilon]^{-1}~=~
\frac{\partial}{\partial\w} \ln [\w^{+} - \epsilon_{d} - \Sigma(\w) -\epsilon ]+
[\w^{+} - \epsilon_{d} - \Sigma(\w) -\epsilon ]^{-1} \frac{\partial\Sigma(\w)}{\partial \w}$, and noting
that
\begin{equation}
-\tfrac{1}{\pi}\mathrm{Im} 
\int^{0}_{-\infty}d\w~ \frac{\partial}{\partial\w}\ln \left[\w^{+} - \epsilon_{d}^{\pd} -\Sigma(\w) - \epsilon\right]~=~
\theta \left(-[\epsilon_{d}^{*} +\epsilon]\right)
\nonumber
\end{equation}
where $\theta(x)$ is the unit step function (with $\theta(0)=\tfrac{1}{2}$), eqs.\ \ref{eq:4},\ref{eq:5}b yield
\begin{equation}
\label{eq:10}
\tfrac{1}{2}n~=~\int_{-\infty}^{-\epsilon_{d}^{*}} d\epsilon ~ \rho_{0}^{\pd}(\epsilon)~-~\tfrac{1}{\pi}I_{L}
\end{equation}
with
\begin{equation}
\label{eq:11}
I_{L} ~:=~ \mathrm{Im} \int^{0}_{-\infty}d\w ~G(\w) \frac{\partial\Sigma(\w)}{\partial\w}
\end{equation}
the Luttinger integral.\cite{LW1960,*Luttingerf1960FermiSurf,*Luttingerf1961,LTG2014} But for a Fermi liquid, Luttinger's theorem gives $I_{L}=0$ independent of interaction strength (whence the volume of the Fermi surface is unchanged by interactions~\cite{LW1960,*Luttingerf1960FermiSurf,*Luttingerf1961}), and eq.\ \ref{eq:10} reduces to~\cite{dmftgeorgeskotliar}
\begin{equation}
\label{eq:12}
\tfrac{1}{2}n~=~\int_{-\infty}^{-\epsilon_{d}^{*}} d\epsilon ~ \rho_{0}^{\pd}(\epsilon).
\end{equation}
From the perspective of the effective quantum impurity model, this constitutes a Friedel sum 
rule,~\cite{hewsonbook,Langreth1966,LTG2014} where the charge $n$ is determined entirely by the renormalised level 
$\epsilon_{d}^{*}$. 

Consider now the implications of eqs.\ \ref{eq:8},\ref{eq:12}, noting first that at the ph-symmetric point, $\eta =0$, the renormalised level $\epsilon_{d}^{*}=0$ and the charge $n=1$ for all $U$, by symmetry.
In that case eq.\ \ref{eq:8} is just the familiar condition that, throughout the metallic phase, the Kondo resonance in 
$D(\w)$ is pinned at the Fermi level to its non-interacting value $\rho_{0}(0)$. Eq.\ \ref{eq:8} generalises this result to the 
generic ph-asymmetric model; and since $n\leq 1$ for  all $\eta \geq 0$ (Fig.\ \ref{fig:fig1}), eq.\ \ref{eq:12} shows that 
\begin{equation}
\label{eq:12a}
\epsilon_{d}^{*} ~\geq~ 0 ~~~~~~:~\eta \geq 0 .
\end{equation}

But the charge $n \rightarrow 1$ continuously on approaching the MT from the metallic phase $U \rightarrow U_{c}(\eta)-$ in the general asymmetric case; which from eq.\ \ref{eq:12}  arises only for $\epsilon_{d}^{*} =0$. Hence
(eqs.\  \ref{eq:8},\ref{eq:12})
 \begin{equation}
\label{eq:13}
\epsilon_{d}^{*} \rightarrow 0 ~,~~~~ D(0) \rightarrow \rho_{0}^{\pd}(0) ~~~~~:~ U \rightarrow U_{c}(\eta)-
\end{equation}
i.e.\ the renormalised level vanishes continuously as the MT is 
approached from the metallic side; and at $U=U_{c}(\eta)-$, $\epsilon_{d}^{*} =0$ and $D(0)=\rho_{0}(0)$,
just as occurs throughout the metallic phase at ph-symmetry. This indicates an emergent ph-symmetry on approaching the Mott transition from the metal (which will be seen further in secs.\ \ref{subsection:sec4C},\ref{subsection:sec5A}).
Note also that, right up to $U=U_{c}(\eta)-$, the single-particle spectrum precisely at the Fermi level remains finite;
reflecting the fact that the Kondo resonance in $D(\w)$ vanishes `on the spot' as $U=U_{c}(\eta)$ is crossed, with $D(0)$ jumping discontinuously from $\rho_{0}(0) >0$ at $U=U_{c}(\eta)-$ to $0$ at $U=U_{c}(\eta)+$ in the Mott insulator.

 Yet there is of course an elephant in the room. Luttinger's theorem reflects perturbative continuity to the non-interacting limit $U=0$. Eq.\ \ref{eq:12} for $n$ hinges on it, and as such applies only to the metallic Fermi liquid. The Mott insulator by contrast is not adiabatically connected to the non-interacting limit, and the Luttinger theorem $I_{L}=0$ does not in general hold. Were it to do so, then since $n=1$  for the MI, eq.\ \ref{eq:12} would imply $\epsilon_{d}^{*}=0$ 
\emph{throughout} the MI phase for \emph{all} asymmetry $\eta$. This is not however the case (as may be verified in several ways, including numerical calculation via e.g.\ NRG). An alternative strategy must thus be employed; as now considered.


\section{Mott insulator}
\label{section:sec4}

Any lattice-fermion model reduces within DMFT to a self-consistently determined quantum impurity 
model,~\cite{dmftgeorgeskotliar} whence Kondo physics in one form or another is at heart involved -- be it the 
Kondo problem for a metallic host (as for the metallic phase) or for a gapped host (as in the MI). 
In the Fermi liquid metal the standard metallic Kondo effect prevails, quenching completely the electron spin 
degrees of freedom. The ground state is thus non-degenerate with e.g.\ a vanishing $T=0$ entropy. In RG terms 
relevant to the underlying quantum impurity model, the stable fixed point is a Strong Coupling one.

The MI within DMFT is by contrast well known to be characterised by a residual entropy of $k_{B}\ln2$ per site. 
This local double-degeneracy reflects the fact that electron spins are not fully Kondo-quenched, in otherwords that
there is an unquenched local moment per site (denoted by $\tilde{\mu}$). In RG terms, the stable fixed point is 
now a Local Moment one. 

To handle the locally doubly-degenerate MI phase requires a two-self-energy (TSE) description. This we have recently considered in detail,~\cite{LTG2014}  in relation to a broad class of impurity models (which includes the gapped 
Anderson model); and basic ideas and results from which we draw on extensively in the following. Within the TSE 
description, the local propagator $G(\w)$ is expressed as
\begin{equation}
\label{eq:14}
G(\w)~=~ \tfrac{1}{2} \left[ G_{A\sigma}(\w) ~+~ G_{B\sigma}(\w)\right]~.
\end{equation}
Here $G_{A\sigma}(\w)$ refers to the  propagator for local moment $\tilde{\mu} =+\modmutil$, while
$G_{B\sigma}(\w)$ refers to that for $\tilde{\mu} =-\modmutil$. From the invariance of the Hamiltonian under 
spin exchange ($\sigma \leftrightarrow -\sigma$), it follows that $G_{A\sigma}(\w)=G_{B-\sigma}(\w)$; whence 
$G(\w)$ in eq.\ \ref{eq:14} is rotationally invariant, i.e.\ independent of $\sigma$ (as it must be at zero-field). 
Eq.\ \ref{eq:14} may thus be written equivalently as
\begin{equation}
\label{eq:15}
G(\w)~=~ \tfrac{1}{2} \left[ G_{A\uparrow}(\w) ~+~ G_{A\downarrow}(\w)\right],
\end{equation}
enabling us to focus solely on the `$A$'-type propagators, and which form we employ in the following. 
The propagators $G_{A\sigma}(\w)$ are given in terms of the two-self-energies
$\Sigma_{A\sigma}(\w)$ ($=\Sigma_{A\sigma}^{R}(\w)-i\Sigma_{A\sigma}^{I}(\w)$), \emph{viz}
\begin{equation}
\label{eq:16}
G_{A\sigma}(\w) ~=~ \left[\w^{+} -\epsilon_{d}^{\pd} - \Sigma_{A\sigma}(\w) - \tfrac{1}{4}t_{*}^{2} G(\w)
\right]^{-1}~;
\end{equation}
and the local moment $\modmutil$ is given in terms of the $G_{A\sigma}(\w)$ by
\begin{equation}
\label{eq:17}
\modmutil~=~ -\tfrac{1}{\pi}\mathrm{Im} \int^{0}_{-\infty}d\w ~\left[
G_{A\uparrow}(\w) - G_{A\downarrow}(\w)\right]~.
\end{equation}

The following points should be noted here, and will be used in our subsequent analysis (for details see 
ref.\ \onlinecite{LTG2014}):\\
\noindent
(i) As above, the local double-degeneracy of the MI phase  reflects an unquenched local moment.
That local degeneracy can be removed by applying a \emph{local} magnetic field $h$ to any given site (i.e.\ to the impurity 
itself in the effective quantum impurity model); via a local field term in the Hamiltonian, 
$-(\hat{n}_{i\uparrow}-\hat{n}_{i\downarrow})h$.
The local magnetisation $m(h) =\langle \hat{n}_{i\uparrow}-\hat{n}_{i\downarrow}\rangle$ 
is then characteristically discontinuous across $h=0$; with $m(h=0+) =+\modmutil$ and $m(h=0-) =-\modmutil$ 
(giving the physical origin of the `$A$' and `$B$'-type zero-field propagators considered above). \\
\noindent 
(ii) For the $h=0$ case of interest to us here, any given site has local moment $\tilde{\mu}=\pm \modmutil$ with equal probability; whence eq.\ \ref{eq:14} for the averaged local propagator has the obvious statistical interpretation.
By the same token, the local Feenberg self-energy  $S(\w)=\sum_{j} t^{2}G_{jj;\sigma}(\w)$ entering eq.\ \ref{eq:16}
indeed becomes $S(\w)=\tfrac{1}{4}t_{*}^{2}G(\w)$ (recall $t= t_{*}/(2\sqrt{Z_{c}})$), since the
$Z_{c}\rightarrow \infty$ sites $j$ which are nearest neighbours to any given (`impurity') site  are equally probably 
`$A$'-type ($\tilde{\mu} = + \modmutil$) as `$B$'-type ($\tilde{\mu} = -\modmutil$).\\
\noindent (iii) As detailed in ref.\ \onlinecite{LTG2014}, it is the self-energies $\Sigma_{A\sigma}(\omega)$ entering eq.\ \ref{eq:16} for the local moment (i.e.\ MI) phase that are directly calculable from many-body perturbation theory, as functional derivatives of a Luttinger-Ward functional. In consequence, a Luttinger theorem holds~\cite{LTG2014} for the
two-self-energies and their associated propagators, \emph{viz}
\begin{equation}
\label{eq:18}
I_{L_{A\sigma}} ~:=~ \mathrm{Im} \int^{0}_{-\infty}d\w ~G_{A\sigma}(\w) \frac{\partial\Sigma_{A\sigma}(\w)}{\partial\w} ~=~0
\end{equation}
(holding separately for each $\sigma$, and for any $U$ in the MI); this result will be employed centrally
in the following.

 By contrast, the standard Luttinger theorem $I_{L}=0$ -- given (see eq.\ \ref{eq:11}) in terms of $G(\w)$ and the conventional single self-energy $\Sigma(\w)$ of standard field theory --
does \emph{not} hold in the MI phase (we determine it explicitly in sec.\ \ref{subsection:sec4E}). 
$\Sigma(\w)$ nevertheless remains defined in the MI phase just as in eqs.\ \ref{eq:5}b or \ref{eq:6}.
Direct comparison between eqs.\ \ref{eq:15},\ref{eq:16} and eq.\ \ref{eq:6} then provides the relation
between the two-self-energies $\{\Sigma_{A\sigma}(\w)\}$ and $\Sigma(\w)$ in the MI phase:
\begin{equation}
\label{eq:19}
\begin{split}
\Sigma(\w) ~&=~ \tfrac{1}{2}\left[\Sigma_{A\uparrow}(\w) + \Sigma_{A\downarrow}(\w)\right] 
\\
+~& \frac{\left[\tfrac{1}{2}\left(\Sigma_{A\uparrow}(\w) - \Sigma_{A\downarrow}(\w)\right)\right]^{2}}{\Big(\w^{+}-\epsilon_{d}^{\pd} -\tfrac{1}{4}t^{2}_{*}G(\w)\Big)~-~\tfrac{1}{2} \left[\Sigma_{A\uparrow}(\w) + \Sigma_{A\downarrow}(\w)\right]}
\end{split}
\end{equation}

We emphasise here that the results above for the MI phase are exact. A TSE description, and the notion of well-formed local moments, is central also to the local moment approach (LMA); which provides a rather successful description of
metallic,~\cite{LET,*MTGLMA_asym,*nigelscalspec}$^{,}$\cite{NLDhEPL,*nigelfield,*nigeltherm,*MRGSUN}
pseudogapped,~\cite{MTG_SPAIM,*nrglmacomp,*MTG_APAIM,*MTGSPAIMepl,*mattgarethPAIM}
and gapped~\cite{mrggaplma,*mrggappt} impurity models, as  well as correlated lattice fermion models within 
DMFT.~\cite{fnrefLMAPAM,Kauch2012} The LMA is however approximate in general, in contrast to the present work. 
We add further that both $\Sigma_{A\sigma}(\w)$ and $\Sigma(\w)$ can be calculated
using NRG, as discussed in ref.\ \onlinecite{LTG2014}.\\

It is also physically instructive to comment on the fact that eq.\ \ref{eq:5}b for $G(\w)$
holds in both the metallic and the MI phases. Its origin in the former case is usually viewed as reflecting 
the translational invariance of electronic states appropriate to the non-degenerate metal,
\emph{viz} 
\begin{equation}
\label{eq:20}
\begin{split}
G(\w) ~&\equiv ~ N^{-1}\sum_{\mathbf{k}} \frac{1}{\w^{+} - \epsilon_{d}^{\pd} -\Sigma(\w) - \epsilon_{\mathbf{k}}}
\\
&=~ \int^{\infty}_{-\infty}d\epsilon~ \frac{\rho_{0}(\epsilon)}{\w^{+} - \epsilon_{d}^{\pd} -\Sigma(\w) -\epsilon}
\end{split}
\end{equation}
with a purely local ($\mathbf{k}$-independent) self-energy 
(and $\rho_{0}(\epsilon) =N^{-1}\sum_{\mathbf{k}} \delta(\epsilon -\epsilon_{\mathbf{k}})$).
The locality of the self-energy is of course intrinsic to DMFT, regardless of the phase. But translational invariance
of electronic states is not, and the `spin-disorder' inherent to the Mott insulator means strictly that 
it is not translationally invariant.
Eq.\ \ref{eq:5}b/\ref{eq:20} nevertheless holds also in the MI, because the Feenberg self-energy $S(\w)$ is the 
same function of $G(\w)$ in both the metallic and MI phases (being the same function of $G$ as it is of the non-interacting propagator for $U=0$); with $S \equiv S(G(\w))$ thus given  by
\begin{equation}
\label{eq:21}
G(\w)~=~ \int_{-\infty}^{\infty}d\epsilon ~ \frac{\rho_{0}(\epsilon)}{S(\w) + \frac{1}{G(\w)} -\epsilon}
\end{equation}
(e.g.\ with $\rho_{0}(\epsilon)$ from eq.\ \ref{eq:2}, eq.\ \ref{eq:21} gives $S(\w)= \tfrac{1}{4}t_{*}^{2}G(\w)$).
But the single self-energy in the MI phase is defined by
$G(\w) = [\w^{+} -\epsilon_{d} -\Sigma(\w) - S(\w)]^{-1}$ (as in eq.\ \ref{eq:6}); whence
$\tfrac{1}{G(\w)}+S(\w) = \w^{+} -\epsilon_{d} -\Sigma(\w)$, and eq.\ \ref{eq:21} thus yields
eq.\ \ref{eq:5}b/\ref{eq:20}. That this holds in the MI phase reflects physically the fact that the 
conventional single self-energy in this phase is by construction that of an effective medium 
(or CPA) description, that is perforce translationally invariant.~\cite{fnepsilondecomp}


\subsection{Local charge}
\label{subsection:sec4A}

Using the above we first obtain a general result (eq.\ \ref{eq:26} below) for the charge $n$ in the MI phase,  
in terms of renormalised levels associated with the two-self-energies; with $n$ given as ever 
by eq.\ \ref{eq:4} (and $n=1$ throughout the MI). With the obvious identity
\begin{equation}
\label{eq:22}
\begin{split}
G_{A\sigma}(\w) ~=&~ \frac{\partial}{\partial\w} \ln\left[\w^{+}-\epsilon_{d}^{\pd} - \Sigma_{A\sigma}(\w) - 
\tfrac{1}{4}t_{*}^{2}G(\w)\right] \\
&+ G_{A\sigma}(\w) \frac{\partial\Sigma_{A\sigma}(\w)}{\partial\w} ~+~
G_{A\sigma}(\w)  \tfrac{1}{4}t_{*}^{2} \frac{\partial G(\w)}{\partial\w},
\end{split}
\end{equation}
eqs.\ \ref{eq:15},\ref{eq:16} and \ref{eq:4} give
\begin{equation}
\begin{split}
n = & \sum_{\sigma}\tfrac{(-1)}{\pi}\mathrm{Im} \int_{-\infty}^{0}d\w ~
\frac{\partial}{\partial\w} \ln\left[\w^{+}-\epsilon_{d}^{\pd} - \Sigma_{A\sigma}(\w) - \tfrac{1}{4}t_{*}^{2}G(\w)\right] 
\\
&+~\tfrac{1}{4}t_{*}^{2}~ \sum_{\sigma} \tfrac{(-1)}{\pi}\mathrm{Im} \int_{-\infty}^{0}d\w ~
G_{A\sigma}(\w)\frac{\partial G(\w)}{\partial\w}
\end{split}
\nonumber
\end{equation}
where the Luttinger theorem $I_{L_{A\sigma}}=0$ (eq.\ \ref{eq:18}) has been used; or equivalently
\begin{equation}
\label{eq:23}
\begin{split}
n =&\sum_{\sigma}\tfrac{(-1)}{\pi}\mathrm{Im} \int_{-\infty}^{0}d\w 
\frac{\partial}{\partial\w} \ln\left[\w^{+}-\epsilon_{d}^{\pd} - \Sigma_{A\sigma}(\w) - \tfrac{1}{4}t_{*}^{2}G(\w)\right] 
\\
&~+\tfrac{1}{2}t_{*}^{2} D(0) G^{R}(0)
\end{split}
\end{equation}
(using eq.\ \ref{eq:15}, and $G(\w)=G^{R}(\w)-i\pi D(\w)$ with $G(\w =-\infty)=0$).
The first term in eq.\ \ref{eq:23} can be determined  in terms of the interaction-renormalised levels 
$\epsilon_{d\sigma}^{*}$ associated with the two-self-energies, given (\emph{cf} eq.\ \ref{eq:9}) by
\begin{equation}
\label{eq:24}
\epsilon_{d\sigma}^{*}~=~ \epsilon_{d}^{\pd}~+~\Sigma_{A\sigma}^{R}(0);
\end{equation}
or equivalently in terms of the `full' renormalised levels
\begin{equation}
\label{eq:25}
\tilde{\epsilon}_{d\sigma}^{*}~=~ \epsilon_{d\sigma}^{*}+\tfrac{1}{4}t_{*}^{2}G^{R}(0)
\end{equation}
(on recognising $\tfrac{1}{4}t_{*}^{2}G(\w)$ as the effective hybridization function `$\Gamma(\w)$' for the 
effective self-consistent impurity model, the $\tilde{\epsilon}_{d\sigma}^{*}$
are effective levels renormalised by both interactions and the hybridization~\cite{LTG2014}). 
Note that since $G(\w)$ is determined self-consistently, $G^{R}(0)$, and hence the $\tilde{\epsilon}_{d\sigma}^{*}$, depend solely on the $\epsilon_{d\sigma}^{*}$ (we return to the matter in secs.\ \ref{subsection:sec4C},\ref{subsection:sec4D}).
Evaluating the first term in eq.\ \ref{eq:23} (using $\Sigma_{A\sigma}^{I}(0)=0$ for the gapped MI) gives
\begin{equation}
n= \sum_{\sigma}\left[
1- \tfrac{1}{\pi}\mathrm{tan}^{-1}\left(
\frac{0^{+} + \tfrac{\pi}{4}t_{*}^{2}D(0)}{-\tilde{\epsilon}_{d\sigma}^{*}}\right) 
\right]
+\tfrac{1}{2}t_{*}^{2} D(0) G^{R}(0)
\nonumber
\end{equation}
(where the arctan $\in [0,\pi]$).
But for the MI, $D(0)=0$ and $n=1$, whence
\begin{equation}
\label{eq:26}
n~=~1~=~ \sum_{\sigma}~
\theta \left(
-\tilde{\epsilon}_{d\sigma}^{*}\right).
\end{equation} 
The Mott insulator thus arises over any interval in which $\tilde{\epsilon}_{d\sigma}^{*}>0$ for one spin, $\sigma$  
(which we show below to be $\sigma =\downarrow$) and $\tilde{\epsilon}_{d-\sigma}^{*}<0$ for spin $-\sigma$.
Importantly,  
and bearing in mind the discussion of sec.\ \ref{section:sec3},
note that the mere existence of a  \emph{range} of 
$\tilde{\epsilon}_{d\sigma}^{*}$ over which the MI can occur, is a direct consequence of the two-self-energy 
description that reflects the inherent degeneracy of the MI phase.


\subsection{Local moment}
\label{subsection:sec4B}

 We turn now to the local moment $\modmutil$. Using again the identity eq.\ \ref{eq:22}, eq.\ \ref{eq:17} gives
\begin{equation}
\begin{split}
~~~~&\modmutil ~=~ 
\\
& \sum_{\sigma} \sigma \tfrac{(-1)}{\pi}\mathrm{Im} \int_{-\infty}^{0}d\w 
\frac{\partial}{\partial\w} \ln\left[\w^{+}-\epsilon_{d}^{\pd} - \Sigma_{A\sigma}(\w) - \tfrac{1}{4}t_{*}^{2}G(\w)\right] 
\\
&+~\tfrac{1}{4}t_{*}^{2}~ \sum_{\sigma} \sigma ~ \tfrac{(-1)}{\pi}\mathrm{Im} \int_{-\infty}^{0}d\w ~
G_{A\sigma}(\w)\frac{\partial G(\w)}{\partial\w}
\end{split}
\nonumber
\end{equation}
(where the Luttinger theorem eq.\ \ref{eq:18} is again used). Proceding analogously to the calculation above, this 
is readily shown to reduce to
\begin{equation}
\label{eq:27}
\begin{split}
~~\modmutil ~=&~\theta \left(\tilde{\epsilon}_{d\downarrow}^{*}\right) ~-~
\theta \left(\tilde{\epsilon}_{d\uparrow}^{*}\right) \\ 
-&~
\tfrac{1}{4}t_{*}^{2}\int_{-\infty}^{0}d\w 
\left[D_{A\downarrow}(\w)\frac{\partial G_{A\uparrow}^{R}(\w)}{\partial\w} +
G_{A\downarrow}^{R}(\w)\frac{\partial D_{A\uparrow}(\w)}{\partial\w}
\right]
\end{split}
\end{equation}
(where the step functions arise from the first term in the previous equation).
Eq.\ \ref{eq:27} is the the $\modmutil$-analogue of eq.\ \ref{eq:26} for $n$. Now consider its implications.

 Recall that $\modmutil >0$, with $\modmutil \in (0,1)$. Deep in the MI ($U \gg U_{c}(\eta)$) a standard perturbative calculation in $t_{*}/U$ gives $\modmutil$ to leading (second) order in $t_{*}/U$. The result is 
\begin{equation}
\label{eq:28}
\modmutil~\overset{U/t_{*} \gg 1}{\sim}~ 1 ~-~ \tfrac{1}{4}\left(\tfrac{t_{*}}{U}\right)^{2}
\end{equation}
(holding for any lattice, and all ph-asymmetry $\eta$), with corrections ${\cal{O}}([t_{*}/U]^{4})$. As trivially confirmed, the leading correction is also precisely what arises from the final term of eq.\ \ref{eq:27}, 
on employing the limiting $t_{*} =0$ (i.e.\ atomic or `Hubbard atom' limit) propagators therein;
\emph{viz}~\cite{LTG2014}
$G_{A\uparrow}(\w) = [\w^{+} -\epsilon_{d}]^{-1}$ and $G_{A\downarrow}(\w) = [\w^{+} -\epsilon_{d} -U]^{-1}$.
Hence, comparing eqs.\ \ref{eq:28},\ref{eq:27}, it follows that
\begin{equation}
\label{eq:29}
\tilde{\epsilon}_{d\downarrow}^{*}>0, ~~~~~~~~
\tilde{\epsilon}_{d\uparrow}^{*}<0
\end{equation}
(such that the first step function in eq.\ \ref{eq:27} is unity, while the second vanishes).
But $\tilde{\epsilon}_{d\downarrow}^{*}$ cannot change sign throughout the
MI phase, otherwise (from eq.\ \ref{eq:27}) $\modmutil$ would  decrease by unity (and thus 
contradict $\modmutil >0$). 
Hence $\tilde{\epsilon}_{d\downarrow}^{*}>0$ for \emph{all}
$U>U_{c}(\eta)$ throughout the MI phase (and likewise $\tilde{\epsilon}_{d\uparrow}^{*}<0$, 
as required by eq.\ \ref{eq:26} for $n=1$).
Equivalently, from eq.\ \ref{eq:29},
$\tilde{\epsilon}_{d\downarrow}^{*} - \tilde{\epsilon}_{d\uparrow}^{*}>0$, i.e.\
(from eq.\ \ref{eq:25}) $\epsilon_{d\downarrow}^{*} - \epsilon_{d\uparrow}^{*}>0$, whence
\begin{equation}
\label{eq:30}
\epsilon_{d\downarrow}^{*} > \epsilon_{d\uparrow}^{*}
\end{equation}
for all $U>U_{c}(\eta)$.

The MI is thus characterised by a charge $n=1$, and a local moment $\modmutil$ which reflects 
the local double-degeneracy of the MI; but with $\modmutil <1$, reflecting the fact that the phase does not consist
simply of free spins. An obvious question then arises: how does $\modmutil$ behave as the MT is approached from the
MI side, $U \rightarrow U_{c}(\eta)+$? The answer is that it tends to a finite value, vanishing discontinuously
on crossing into the Fermi liquid metal. 
The reason is physically obvious. The critical $U_{c}(\eta)$ for the MT is a `$U_{c2}$', such
that for $U<U_{c}(\eta)$ the non-degenerate metal is the ground state. A Mott insulating solution nevertheless exists
in an interval $U_{c1}(\eta) <U<U_{c}(\eta)$ (but with a higher energy than the metallic ground state), 
and is continuous across $U=U_{c}(\eta)$. Since the local moment $\modmutil$ for the insulating solution is non-vanishing,
$\modmutil$ will thus remain finite as the MT is approached from the MI, $U\rightarrow U_{c}(\eta)+$.

There are two further consequences of the fact that the moment remains finite down to $U=U_{c}(\eta)+$
(following, as in ref.\ \onlinecite{LTG2014}, because the underlying impurity model in the MI is that for a 
degenerate local moment phase): \\
\noindent (a) The finite-temperature, zero-field local spin susceptibility in response to a field $h$
applied locally to site $i$, $\chi_{i}^{\pd}(T; h=0)= (\partial m(T,h)/\partial h)_{h=0}$ (with
$m(T,h)=\langle\hat{n}_{i\uparrow}-\hat{n}_{i\downarrow}\rangle$), has  the leading $T\rightarrow 0$ behaviour 
 \begin{equation}
\label{eq:31}
\underset{T\rightarrow 0}{\mathrm{lim}}~T\chi_{i}^{\pd}(T; h=0)~=~ \modmutil^{2}~,
\end{equation}
i.e.\ the expected Curie form, with a coefficient of precisely $\modmutil^{2}$.
This behaviour thus likewise persists down to  $U=U_{c}(\eta)+$.\\
\noindent (b) The $T=0$ local spin susceptibility,
$\chi_{i}^{\pd}(T=0; h\rightarrow 0) =(\partial m(0,h)/\partial h )_{h\rightarrow 0}$, is of form 
\begin{equation}
\label{eq:32}
\chi_{i}^{\pd}(T=0; h\rightarrow 0) ~=~2\modmutil\delta(h) +\chi_{i}^{\pd}(T=0; h=0+)
\end{equation}
where $\chi_{i}^{\pd}(T=0; h=0+)$ itself
remains finite as $U$$\rightarrow U_{c}+$ (again reflecting the finite $\modmutil$ 
throughout the MI).\\

\begin{figure}
\includegraphics{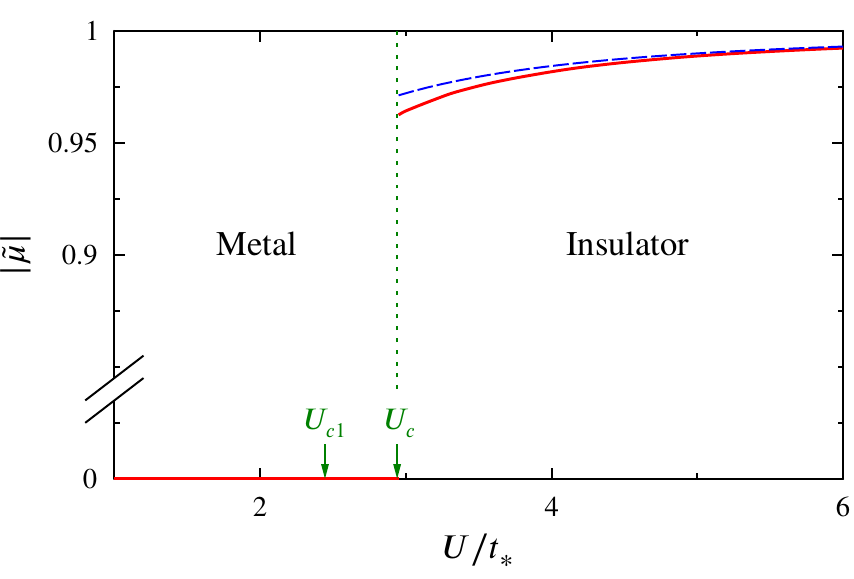}
\caption{\label{fig:fig2} 
NRG calculations (solid line) for the moment $\modmutil$ \emph{vs} $U/t_{*}$. Shown for ph-symmetry $\eta =0$, with 
$U_{c}(0)/t_{*} \simeq 2.95$ indicated. $\modmutil$ is non-zero only in the MI phase, and remains finite down to $U=U_{c}(0)+$.
Dashed line shows the asymptotic behaviour eq.\ \ref{eq:28}.  $U_{c1}(0)/t_{*} \simeq 2.45$ is also indicated.
}
\end{figure}

The theory above is fully supported by NRG calculations. Fig.\ \ref{fig:fig2} shows NRG results for the local moment
$\modmutil$ \emph{vs} $U/t_{*}$, at ph-symmetry $\eta =0$. The moment indeed remains finite through the MI, right down to 
$U_{c}(\eta =0)/t_{*} \simeq 2.95$. It is in fact close to the asymptotic behaviour eq.\ \ref{eq:28} throughout the MI; 
reflecting the fact that $U_{c}(\eta =0)/t_{*}$ appreciably exceeds unity, whence local moments are well-developed in the insulator. The $U_{c1}(0)/t_{*} \simeq 2.45$, down to which an insulating solution exists, is also indicated. For this solution (which is not of course the ground state below $U_{c}(0)$), we find that a non-zero moment $\modmutil$ in fact persists right down to $U=U_{c1}(0)$.
The same behaviour as fig.\ \ref{fig:fig2} is found for any $\eta \in [0,1)$ (where the MI phase exists); and since
$U_{c}(\eta)$ for $\eta >0$ exceeds $U_{c}(0)$ (fig.\ \ref{fig:fig1}), local moments in the MI are even more strongly developed than for $\eta =0$, with the leading asymptotics for $\modmutil$ (eq.\ \ref{eq:28}) accordingly followed increasingly closely throughout the MI.

\begin{figure}
\includegraphics{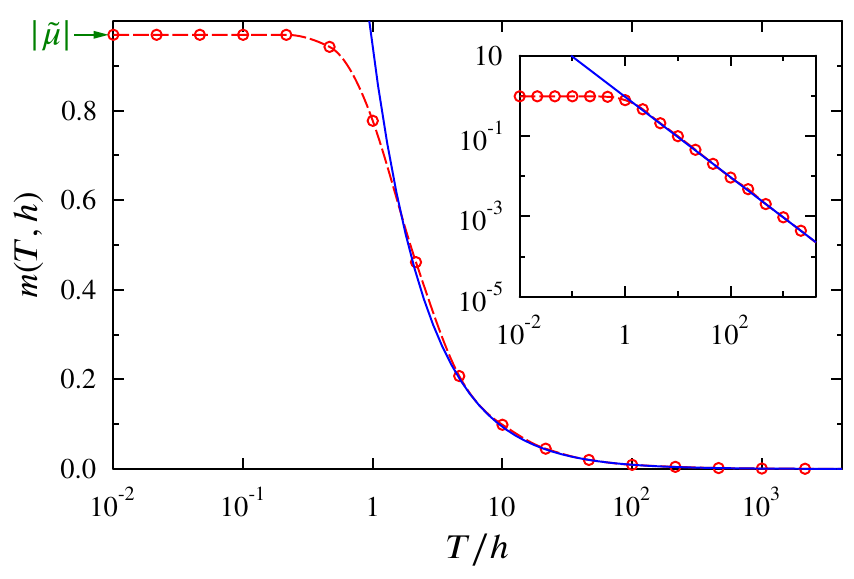}
\caption{\label{fig:fig3} 
NRG results for the $\eta =0$ MI, with $U=3.2t_{*}$ and a tiny fixed local field $h/t_{*}=10^{-5}$.  
Open circles (with dashed line as guide to eye) show the $T$-dependence of the local magnetisation 
$m(T,h)=\langle\hat{n}_{i\uparrow}-\hat{n}_{i\downarrow}\rangle$ \emph{vs} $T/h$ (on a log-scale).
The $T=0$ local moment $\modmutil$ is indicated by an arrow.
Solid line shows $\modmutil^{2}h/T$ with the local moment $\modmutil$ from fig.\ \ref{fig:fig2}, corresponding to
the behaviour eq.\ \ref{eq:31}.
\emph{Inset:} Same results, with both axes on a log-scale. For discussion, see text.
}
\end{figure}

NRG results (again for $\eta =0$) are shown in fig.\ \ref{fig:fig3} for the $T$-dependence of the local magnetisation $m(T,h)$ in the MI ($U=3.2t_{*} >U_{c}(0)$), for a tiny fixed local field $h/t_{*} =10^{-5}$; with $m(T,h)$ plotted 
as a function of $T/h$.
These illustrate nicely the non-commuting order of limits, $T\rightarrow 0$ and $h\rightarrow 0$, that characterises the MI 
(local moment) phase.~\cite{LTG2014} For $T\rightarrow 0$, followed by $h\rightarrow 0+$, the local magnetisation reduces to the local moment $\modmutil$, which is marked on fig.\ \ref{fig:fig3} (and is indistinguishable from $m(T=0, h)$ for the tiny field considered). For the reverse order by contrast, eq.\  \ref{eq:31} gives $m \sim \modmutil^{2}h/T$. This behaviour is indeed seen to arise in fig.\ \ref{fig:fig3} (in practice for $T/h \gtrsim 1$): the solid line shows $\modmutil^{2}h/T$ with the local moment $\modmutil$ taken from fig.\ \ref{fig:fig2}.

\subsection{Renormalised levels $\epsilon_{d\sigma}^{*}$ and $\tilde{\epsilon}_{d\sigma}^{*}$}
\label{subsection:sec4C}

Here we consider the renormalised levels $\epsilon_{d\sigma}^{*}$ and $\tilde{\epsilon}_{d\sigma}^{*}$, given by 
eqs.\ \ref{eq:24},\ref{eq:25} in terms of the two-self-energies (and characteristic of the Mott insulator in analogy 
to the way that $\epsilon_{d}^{*}$ (eq.\ \ref{eq:9}) is characteristic of the metal, sec.\ \ref{section:sec3}).

Note that under a ph-transformation $c_{j\sigma}\leftrightarrow (-1)^{j}c_{j-\sigma}^{\dagger}$, 
it is readily shown that
\begin{equation}
\label{eq:33}
\begin{split}
\Sigma_{A\sigma}^{R}(\w; \eta) ~=&~ U-\Sigma_{A-\sigma}^{R}(-\w; -\eta) 
\\
G^{R}(\w; \eta) ~=&~-G^{R}(-\w; -\eta)
\end{split}
\end{equation}
(with the $\eta$-dependence temporarily explicit), and in consequence the renormalised levels satisfy
\begin{equation}
\label{eq:34}
\epsilon_{d\sigma}^{*}(\eta)~=~ - \epsilon_{d-\sigma}^{*}(-\eta), ~~
\tilde{\epsilon}_{d\sigma}^{*}(\eta)~=~ - \tilde{\epsilon}_{d-\sigma}^{*}(-\eta)~.
\end{equation}
Given this, it is convenient to define
\begin{equation}
\label{eq:35}
\epsilon_{*} ~=~ \tfrac{1}{2}\left(\epsilon_{d\downarrow}^{*} + \epsilon_{d\uparrow}^{*}\right) ~~~~~~
\delta\epsilon_{*} ~=~ \tfrac{1}{2}\left(\epsilon_{d\downarrow}^{*} - \epsilon_{d\uparrow}^{*}\right)
\end{equation}
which are thus respectively odd and even in $\eta$. For use in sec.\ \ref{subsection:sec4D}, we also define 
\begin{equation}
\label{eq:36}
\gamma_{\sigma}~=~ \left(
\frac{\partial\Sigma_{A\sigma}^{R}(\w)}{\partial\w}
\right)_{\w =0}
\end{equation}
which in physical terms is related to the quasiparticle weight 
$Z_{\sigma} = [1-(\partial\Sigma_{A\sigma}^{R}(\w)/\partial\w)_{0}]^{-1}$ for
the self-energy $\Sigma_{A\sigma}(\w)$ by $\gamma_{\sigma} = [1-Z_{\sigma}^{-1}]$; and from eq.\ \ref{eq:33} satisfies
\begin{equation}
\label{eq:37}
\gamma_{\sigma}(\eta)~=~\gamma_{-\sigma}(-\eta)~.
\end{equation}

$\tilde{\epsilon}_{d\sigma}^{*}$ and $\epsilon_{d\sigma}^{*}$ differ (eqs.\ \ref{eq:24},\ref{eq:25}) solely by the
`hybridization' contribution of $\tfrac{1}{4}t_{*}^{2}G^{R}(0)$ to the former; and $G^{R}(0)$ depends solely on the 
$\{\epsilon_{d\sigma}^{*}\}$ -- or equivalently upon $\epsilon_{*}$ and $\delta\epsilon_{*}$ 
(eq.\ \ref{eq:35}). From the basic self-consistency equations \ref{eq:15},\ref{eq:16}, $G(0)$ is given from solution of
\begin{equation}
\label{eq:38}
\Big(\left[\epsilon_{*} +\tfrac{1}{4}t_{*}^{2}G(0)\right]^{2} -\left[\delta\epsilon_{*}\right]^{2}\Big) G(0)
~=~ - \left[\epsilon_{*} +\tfrac{1}{4}t_{*}^{2}G(0)\right].
\end{equation}

The above equations refer generally to the MI phase. But they also hold equally in the metallic phase, simply on dropping the
$\sigma$-labels in $\epsilon_{d\sigma}^{*}$, such that (eq.\ \ref{eq:35}) $\epsilon_{*} \equiv \epsilon_{d}^{*}$
and $\delta\epsilon_{*} =0$. So consider briefly the metal. In this case, eq.\ \ref{eq:38} gives 
$[\epsilon_{d}^{*} +\tfrac{1}{4}t_{*}^{2}G(0)]^{2}G(0) = -[\epsilon_{d}^{*} +\tfrac{1}{4}t_{*}^{2}G(0)]$,
to which the physical solution is 
$[\epsilon_{d}^{*} +\tfrac{1}{4}t_{*}^{2}G(0)]G(0) = -1$.~\cite{fnphyssoln1}
But as shown in sec.\ \ref{section:sec3} (eq.\ \ref{eq:13}), $\epsilon_{d}^{*} \rightarrow 0$ as $U \rightarrow U_{c}(\eta)-$,
whence $\tfrac{1}{4}t_{*}^{2}G(0)^{2}= -1$ for $U=U_{c}(\eta)-$. $G(0)$ is thus pure imaginary 
(recovering $D(0)=\rho_{0}(0)$ using eq.\ \ref{eq:2}, as in eq.\ \ref{eq:13}), with
\begin{equation}
\label{eq:39}
G^{R}(\w=0)~=~ 0      ~~~~~~:~ U=U_{c}(\eta)-.
\end{equation}
As for $\epsilon_{d}^{*}$, the `full' renormalized level
$\tilde{\epsilon}_{d}^{*} =\epsilon_{d}^{*} + \tfrac{1}{4}t_{*}^{2}G^{R}(0)$ thus vanishes
as the Mott transition is approached from the metallic side, $U\rightarrow U_{c}(\eta)-$.
Recall moreover that eq.\ \ref{eq:39} holds for \emph{any} asymmetry $\eta \in [0,1)$; which again shows the emergent
ph-symmetry (sec.\ \ref{section:sec3}) on approaching the transition from the metal (noting from
eq.\ \ref{eq:33} that $G^{R}(0) =0$ at the ph-symmetric point $\eta =0$).

Now return to the MI phase. For the particular case of ph-symmetry $\eta=0$, eq.\ \ref{eq:34}  gives             
\begin{equation}
\label{eq:40}
\epsilon_{d\uparrow}^{*} ~=~-\epsilon_{d\downarrow}^{*},
~~~~\tilde{\epsilon}_{d\uparrow}^{*} ~=~-\tilde{\epsilon}_{d\downarrow}^{*}
~~~~~~:~ \eta=0
\end{equation}
such that, throughout the MI, 
$\epsilon_{*}=\tfrac{1}{2}(\epsilon_{d\downarrow}^{*} + \epsilon_{d\uparrow}^{*}) =0=\tfrac{1}{2}(\tilde{\epsilon}_{d\downarrow}^{*} + \tilde{\epsilon}_{d\uparrow}^{*})$;
and similarly (eq.\ \ref{eq:33}), $G^{R}(0) = 0$ for \emph{all} $U$.
An obvious question is: how do the renormalised levels $\epsilon_{d\sigma}^{*}$ and $\tilde{\epsilon}_{d\sigma}^{*}$ in general
behave as the MT is approached, $U\rightarrow U_{c}(\eta)+$? As for the local moment $\modmutil$ considered in 
sec.\ \ref{subsection:sec4B}, the answer is that they tend to finite values, and the basic reason is 
again that given in sec.\ \ref{subsection:sec4B}:
on approaching $U_{c}(\eta)$ from the MI, the insulating solution \emph{itself} is continuous across 
$U_{c}(\eta) \equiv U_{c2}(\eta)$, persists down to $U = U_{c1}(\eta)$, and does not `know' about the metallic solution at 
$U_{c}(\eta)-$. Since the bounds on the $\tilde{\epsilon}_{d\sigma}^{*}$ and $\epsilon_{d\sigma}^{*}$ 
established in eqs.\ \ref{eq:29},\ref{eq:30} hold equally for the insulating solution down to $U=U_{c1}(\eta)$, 
all renormalised levels are thus finite at the transition $U=U_{c}(\eta)+$, for any $\eta$.~\\

\begin{figure}
\includegraphics{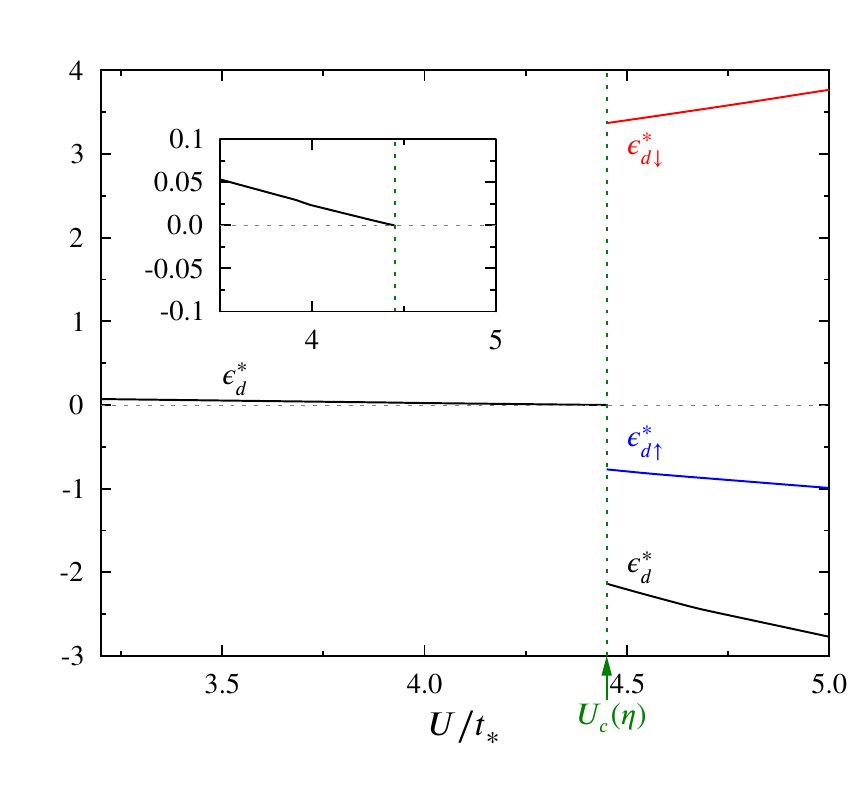}
\caption{\label{fig:fig4} 
NRG results for renormalised levels, shown \emph{vs} $U/t_{*}$ for asymmetry $\eta =0.5$ ($U_{c}(\eta)/t_{*} \simeq 4.48$),
with $t_{*}\equiv 1$ taken on the vertical axis. On approaching the transition from the Mott insulator, 
$\epsilon_{d\downarrow}^{*}$ and 
$\epsilon_{d\uparrow}^{*}$ indeed remain finite as $U\rightarrow U_{c}(\eta)+$. On approaching from the metal
by contrast, $\epsilon_{d}^{*}$ (eq.\ \ref{eq:9}) vanishes linearly in $U_{c}(\eta)-U$ (see inset).
NRG results for $\epsilon_{d}^{*}$ in the MI (again defined by eq.\ \ref{eq:9}), are also plotted. As shown in 
sec.\ \ref{subsection:sec4F}, $\epsilon_{d}^{*}$ necessarily satisfies 
$|\epsilon_{d}^{*}| >t_{*} (\equiv 1)$ throughout the MI, for any asymmetry $\eta$ other than the ph-symmetric point 
$\eta =0$. For $\eta =0$ by contrast, $\epsilon_{d}^{*}$ vanishes throughout \emph{both} phases.
}
\end{figure}

The results above, and those of sec.\ \ref{section:sec3}, are likewise supported by NRG calculations; as illustrated in 
fig.\ \ref{fig:fig4} for fixed asymmetry $\eta =0.5$ ($\epsilon_{d}=-\tfrac{U}{4}$), where the transition occurs at 
$U_{c}(\eta)/t_{*} \simeq 4.48$.  On the metallic side, the renormalised level 
$\epsilon_{d}^{*} =\epsilon_{d}+\Sigma^{R}(0)$ (eq.\ \ref{eq:9}) is non-negative 
(as in eq.\ \ref{eq:12a}). As seen clearly from the inset to fig.\ \ref{fig:fig4}, it indeed vanishes as 
$U\rightarrow U_{c}(\eta)-$ (eq.\ \ref{eq:13}), and does so with an exponent of unity, 
\begin{equation}
\label{eq:41}
\epsilon_{d}^{*} ~\overset{U\rightarrow U_{c}(\eta)-}{\propto}~ \left(U_{c}-U\right)
\end{equation}
(which behaviour is found to be generic).
From eq.\ \ref{eq:12}, $1-n \sim 2\rho_{0}(0)\epsilon_{d}^{*}$ as $\epsilon_{d}^{*}\rightarrow 0$, 
whence $1-n$ likewise vanishes linearly as the transition is approached from the metal, 
\begin{equation}
\label{eq:42}
1-n ~\overset{U\rightarrow U_{c}(\eta)-}{\propto}~ \left(U_{c}-U\right).
\end{equation}
On the insulating side by contrast, the renormalised levels $\epsilon_{d\uparrow}^{*}$ and $\epsilon_{d\downarrow}^{*}$ indeed remain finite as $U\rightarrow U_{c}(\eta)+$, and vary near linearly with $U/t_{*}$ in the MI. (The 
$\tilde{\epsilon}_{d\sigma}^{*}$ are close to their $\epsilon_{d\sigma}^{*}$ counterparts, and are omitted from  fig.\ \ref{fig:fig4} for clarity.)


\subsection{Single self-energy, $\Sigma(\w)$}
\label{subsection:sec4D}

With the above in mind, we turn now to the conventional self-energy $\Sigma(\w)$ in the MI, with $\Sigma(\w)$  given by 
eq.\ \ref{eq:19} in terms of the two self-energies $\Sigma_{A\sigma}(\w)$; our particular interest being the low-energy behaviour of $\Sigma(\w)$.

Consider first $U/t_{*} \gg 1$, deep in the MI. Here the spectrum $D(\w)$ consists of Hubbard bands, each of width 
${\cal{O}}(t_{*})$, centred on
$\w_{\pm} = \tfrac{U}{2}\eta \pm \tfrac{U}{2}$ (i.e.\ $\w_{-} = \epsilon_{d}$ and $\w_{+} = \epsilon_{d} +U$),
and with an insulating gap $\Delta \sim  {\cal{O}}(U) \gg t_{*}$. From  Hilbert transformation, $\Sigma_{A\sigma}(\w)$ is given  generally by
\begin{equation}
\label{eq:43}
\Sigma_{A\sigma}(\w) ~=~\Sigma_{A\sigma}^{s}~+~
\int^{\infty}_{-\infty}\frac{d\w_{1}}{\pi} ~\frac{\Sigma_{A\sigma}^{I}(\w_{1})}{\w^{+}-\w_{1}}
\end{equation}
where $\Sigma_{A\sigma}^{s}$ denotes the purely static ($\w$-independent) contribution to the self-energy.
It is given exactly by $\Sigma_{A\sigma}^{s}=U\langle\hat{n}_{i-\sigma}\rangle = \frac{1}{2}U[n-\sigma\modmutil]$, with local
charge $n =1$ throughout the MI and local moment $\modmutil \rightarrow 1$ for $U/t_{*} \gg 1$ 
(eq.\ \ref{eq:28}).
Deep inside the insulating gap, the second (`dynamical') term in eq.\ \ref{eq:43}
gives an asymptotically vanishing contribution to $\Sigma_{A\sigma}(\w)$,
since $\Sigma_{A\sigma}^{I}(\w)$ is non-zero only outside the gap; and by the same argument
$G(\w) =\int^{\infty}_{-\infty}d\w_{1} ~D(\w_{1})/(\w^{+} -\w_{1})$ likewise vanishes asymptotically 
(being ${\cal{O}}(1/U)$ deep in the gap). Hence to leading order, 
 $\Sigma_{A\downarrow}(\w) \equiv \Sigma_{A\downarrow}^{s}= U$,
$\Sigma_{A\uparrow}(\w) \equiv \Sigma_{A\uparrow}^{s}= 0$ and $G(\w) \equiv 0$, and
eq.\ \ref{eq:19} thus gives
\begin{equation}
\label{eq:44}
\Sigma (\w) ~\sim ~ \tfrac{1}{2}U ~+~\frac{\tfrac{1}{4}U^{2}}{\w^{+} - (\epsilon_{d} + \tfrac{1}{2}U)}.
\end{equation}
This result -- which arises as a simple and direct consequence of the underlying two-self-energy description --
is asymptotically exact deep in the insulating gap for $U/t_{*} \gg 1$. It will also be recognised (see e.g.\ 
sec.\ IV of ref.\ \onlinecite{LTG2014}) as the exact self-energy for the $t_{*} =0$ atomic limit for 
\emph{any} asymmetry $\eta$, which is physically natural.
For our present purposes, a key feature of eq.\ \ref{eq:44} is that $\Sigma(\w)$ contains a pole within the
insulating gap, occurring at $\w = \epsilon_{d}+\tfrac{1}{2}U = \tfrac{1}{2}(\w_{+}+\w_{-})$.
We now consider such behaviour more generally throughout the MI, away from the strong coupling limit
$U/t_{*} \gg 1$ and down to the MT occurring at $U=U_{c}(\eta)$. \\

We are interested in the low-$\w$ behaviour of the conventional self-energy $\Sigma(\w)$ in the MI;
`low' here meaning close to the Fermi level $\w =0$ within the gap in $D(\w)$ (where $G(\w)$ 
and the $\Sigma_{A\sigma}(\w)$ are pure real). The leading low-$\w$ behaviour of $\Sigma(\w)$
arises from the final term in eq.\ \ref{eq:19} (so we neglect the first term therein
in the \emph{ff}).
The numerator in eq.\ \ref{eq:19} is recognised  (from eqs.\ \ref{eq:24},\ref{eq:35}) as being 
$\delta\epsilon_{*}^{2} =[\tfrac{1}{2} (\epsilon_{d\downarrow}^{*} -\epsilon_{d\uparrow}^{*})]^{2}$
as $\w \rightarrow 0$, while the denominator can be expanded to linear order in $\w$, to give
\begin{equation}
\label{eq:45}
\Sigma(\w) ~\overset{\w \rightarrow 0}{\sim}~ 
\frac{(\delta\epsilon_{*})^{2}}{i0^{+} -\tfrac{1}{2}(\tilde{\epsilon}_{d\downarrow}^{*} + 
\tilde{\epsilon}_{d\uparrow}^{*}) +\lambda \w}
\end{equation}
where
\begin{equation}
\label{eq:46}
\lambda ~=~1~-~\tfrac{1}{2}\sum_{\sigma}\gamma_{\sigma}~-~
\tfrac{1}{4}t_{*}^{2}\left(\frac{\partial G^{R}(\w)}{\partial\w}\right)_{0}
\end{equation}
and $\gamma_{\sigma}$ is defined in eq.\ \ref{eq:36}.

Eq.\ \ref{eq:45} is general. We now consider it explicitly at, and close to, ph-symmetry 
$\eta =0$. The  behaviour of $G^{R}(0)$ is easily determined by iteration of eq.\ \ref{eq:38}, with
the result   
\begin{equation}
\tfrac{1}{4}t_{*}^{2}G^{R}(0) ~=~ \frac{\epsilon_{*}}{(2\delta\epsilon_{*}/t_{*})^{2}-1} ~+~{\cal{O}}(\epsilon_{*}^{3})
\nonumber
\end{equation}
(such that $G^{R}(0) =0$ for $\epsilon_{*} =0$, as at ph-symmetry).
Note that this holds to leading order in $\epsilon_{*}=\tfrac{1}{2}(\epsilon_{d\downarrow}^{*}+\epsilon_{d\uparrow}^{*})$, 
but for any $\delta\epsilon_{*}=\tfrac{1}{2}(\epsilon_{d\downarrow}^{*}-\epsilon_{d\uparrow}^{*})$.
The `full' renormalized levels 
$\tilde{\epsilon}_{d\sigma}^{*}=\epsilon_{d\sigma}^{*}+\tfrac{1}{4}t_{*}^{2}G^{R}(0)
= \epsilon_{*} -\sigma\delta\epsilon_{*}+\tfrac{1}{4}t_{*}^{2}G^{R}(0)$
(with $\sigma =\pm$ for $\uparrow$$/$$\downarrow$-spins)
then follow, and hence
\begin{equation}
\label{eq:47}
\tfrac{1}{2}\left(\tilde{\epsilon}_{d\downarrow}^{*} + \tilde{\epsilon}_{d\uparrow}^{*}\right)
~=~
\frac{\epsilon_{*}(2\delta\epsilon_{*}/t_{*})^{2}}{(2\delta\epsilon_{*}/t_{*})^{2}-1}~+~{\cal{O}}(\epsilon_{*}^{3}).
\end{equation}


\subsubsection{Particle-hole symmetry}
\label{subsubsection:sec4D1}

Consider first the case of ph-symmetry $\eta=0$,  where
$\tfrac{1}{2}(\tilde{\epsilon}_{d\downarrow}^{*} + \tilde{\epsilon}_{d\uparrow}^{*})=0$ (eq.\ \ref{eq:40}).
From eq.\ \ref{eq:45}, $\Sigma(\w)$ necessarily contains a pole at the Fermi level for all $U >U_{c}(0)$,
\begin{equation}
\label{eq:48}
\Sigma(\w) \overset{\w \rightarrow 0}{\sim} 
\frac{(\delta\epsilon_{*})^{2}}{i0^{+}  +\lambda \w} ~~:~
\delta\epsilon_{*} \equiv  \epsilon_{d\downarrow}^{*} = -\tfrac{1}{2}U +\Sigma_{A\downarrow}^{R}(0).
\end{equation}
As with $G^{R}(0)$, the coefficient $(\partial G^{R}(\w)/\partial\w)_{\w=0}$ entering eq.\ \ref{eq:46} may be determined self-consistently as a function of $\delta\epsilon_{*}$ and $\gamma_{\sigma}$ (eq.\ \ref{eq:36}), using the basic DMFT 
eqs.\ \ref{eq:15},\ref{eq:16}. The analysis is lengthy, but the final result is simple and for $\eta =0$ gives
\begin{equation}
\lambda~=~\frac{(1-\tfrac{1}{2}[\gamma_{\uparrow}+\gamma_{\downarrow}])~
(2\delta\epsilon_{*}/t_{*})^{2}}{(2\delta\epsilon_{*}/t_{*})^{2}-1}
\nonumber
\end{equation}
(with $\gamma_{\sigma}$  independent of $\sigma$ for $\eta =0$, see eq.\ \ref{eq:37}).
Note that
$\Sigma^{I}(\w) \equiv Q_{p}\delta(\w)$, with pole-weight $Q_{p} = \pi(\delta\epsilon_{*})^{2}/|\lambda|$
given by
\begin{equation}
\label{eq:49}
Q_{p}~=~\frac{\pi t_{*}^{2}}{4} \frac{|(2\delta\epsilon_{*}/t_{*})^{2}-1|}{ \left(1+\tfrac{1}{2}[|\gamma_{\uparrow}|+|\gamma_{\downarrow}|]\right)}.
\end{equation}
The existence of an $\w =0$ pole in $\Sigma(\w)$ is a key signature of the MI phase at ph-symmetry.~\cite{dmftgeorgeskotliar} 
Now consider the pole-weight $Q_{p}$. Since $\gamma_{\sigma}=(\partial\Sigma_{A\sigma}^{R}(\w)/\partial\w)_{0}$, it follows
from eq.\ \ref{eq:43} that
$\pi\gamma_{\sigma} =-\int^{\infty}_{-\infty}d\w ~\Sigma_{A\sigma}^{I}(\w)/\w^{2}$ ($<0$ necessarily since
$\Sigma^{I}_{A\sigma}(\w) \ge 0$ by analyticity).
But as $U \rightarrow U_{c}(0)+$ the insulating gap remains finite, and
$\Sigma_{A\sigma}^{I}(\w)$ vanishes within the gap. $\gamma_{\sigma}$ thus remains finite as
$U \rightarrow U_{c}(0)+$ (and $|\gamma_{\sigma}|$ must decrease with increasing gap, i.e.\ with increasing $U$).

What can be said about $(2\delta\epsilon_{*}/t_{*})$ throughout the MI?
First, note that for all $U>U_{c}(\eta)$, 
$\delta\epsilon_{*} =\tfrac{1}{2}(\tilde{\epsilon}_{d\downarrow}^{*}-\tilde{\epsilon}_{d\uparrow}^{*}) >0$ 
necessarily (from eq.\ \ref{eq:29}). Now consider again $U/t_{*} \gg 1$, deep in the MI where $\Sigma(\w)$ is given by
eq.\ \ref{eq:44} above; \emph{viz}
\begin{equation}
\Sigma(\w) ~\overset{\w \rightarrow 0}{\sim}~ \frac{\tfrac{1}{4}U^{2}}{\w+i0^{+}}
~~~~~~:~ U/t_{*} \gg 1 ,
\nonumber
\end{equation}
(which corresponds, \emph{cf} eq.\ \ref{eq:48}, to $\delta\epsilon_{*} = \tfrac{1}{2}U$ and $\lambda =1$).
But the fact that $2\delta\epsilon_{*}/t_{*} = U/t_{*} \gg 1$ for $U/t_{*} \gg 1$ 
means that $2\delta\epsilon_{*}/t_{*}$ must exceed unity for \emph{all} $U >U_{c}(0)$: if it crossed unity for
some $U>U_{c}(0)$ then $Q_{p} =0$ at that point, i.e.\ the pole intrinsic to the MI would vanish. 
The pole could moreover vanish as $U\rightarrow U_{c}(0)+$ only if
$2\delta\epsilon_{*}/t_{*} \rightarrow 1$ (i.e.\ $\epsilon_{d\downarrow}^{*} \rightarrow \tfrac{1}{2}t_{*}$)
as $U\rightarrow U_{c}(0)+$. But there is no reason to expect such a `special' value of $\epsilon_{d\downarrow}^{*}$.
We thus expect the $\w=0$ pole in $\Sigma(\w)$ to remain intact, with non-zero weight, right down to the transition at 
$U_{c}(0)$ --- just as e.g.\ the local moment $\modmutil$ remains finite down to the Mott transition.


\subsubsection{Away from particle-hole symmetry}
\label{subsubsection:sec4D2}

Now consider the situation close to, but away from, ph-symmetry, with $\Sigma(\w)$ at low-energies given by
eq.\ \ref{eq:45}. In this case $\tfrac{1}{2}(\tilde{\epsilon}_{d\downarrow}^{*} + \tilde{\epsilon}_{d\uparrow}^{*})$ 
(which is odd in $\eta$ from eq.\ \ref{eq:34}), is non-vanishing, and given to leading order in $\epsilon_{*}$ by 
eq.\ \ref{eq:47}; while $\lambda$ (eq.\ \ref{eq:46}) is given by
\begin{equation}
\lambda ~=~ \frac{(2\delta\epsilon_{*}/t_{*})^{2}}{(2\delta\epsilon_{*}/t_{*})^{2}-1}~\alpha ~+~
{\cal{O}}(\epsilon_{*}^{2})
\nonumber
\end{equation}
where
\begin{equation}
\alpha ~=~  1 ~-~\tfrac{1}{2}(\gamma_{\uparrow}+\gamma_{\downarrow}) ~-~
\frac{(\gamma_{\uparrow}-\gamma_{\downarrow})}{[(2\delta\epsilon_{*}/t_{*})^{2}-1]} ~\frac{\epsilon_{*}}{\delta\epsilon_{*}}
\nonumber
\end{equation}
(with $\lambda$ and $\alpha$ both even in $\eta$).
From eq.\ \ref{eq:45}, $\Sigma(\w)$ thus has a pole at a non-vanishing energy
$\w = \epsilon_{*}/\alpha$,
\begin{equation}
\label{eq:50}
\Sigma^{I}(\w) ~\overset{\w \rightarrow 0}{\sim}~ Q_{p}~\delta (\w -\tfrac{\epsilon_{*}}{\alpha})
\end{equation}
with pole-weight $Q_{p} = \pi(\delta\epsilon_{*})^{2}/|\lambda|$ given by
\begin{equation}
\label{eq:51}
Q_{p}~=~\frac{\pi t_{*}^{2}}{4} \frac{|(2\delta\epsilon_{*}/t_{*})^{2}-1|}{|\alpha|}.
\end{equation}
$Q_{p} \equiv Q_{p}(\eta)$ is moreover even in $\eta$, so to leading order in asymmetry may be
replaced by its ph-symmetric limit $Q_{p}(\eta=0)$ given by eq.\ \ref{eq:49}, which as argued in 
sec.\ \ref{subsubsection:sec4D1}
remains finite down to the Mott transition.
$\epsilon_{*}$ by contrast is odd in $\eta$.
For sufficiently small asymmetry at least, we are thus guaranteed
a low-energy pole in $\Sigma^{I}(\w)$, which lies in the insulating gap at a non-zero energy away from the Fermi level, and 
persists down to the Mott transition at $U=U_{c}(\eta)+$.


\subsection{Luttinger theorem for $I_{L}$}
\label{subsection:sec4E}

Luttinger's theorem $I_{L}=0$ holds in the Fermi liquid metallic phase (sec.\ \ref{section:sec3}), with the usual Luttinger integral $I_{L}$ given in eq.\ \ref{eq:11} in terms of the self-energy $\Sigma(\w)$. 
A Luttinger theorem $I_{L_{A\sigma}}=0$ also holds~\cite{LTG2014} in the MI (sec.\ \ref{section:sec4}),
with $I_{L_{A\sigma}}$ now given in terms of the \emph{two}-self-energies $\Sigma_{A\sigma}(\w)$ and their associated propagators (eq.\ \ref{eq:18}). However Luttinger's theorem expressed in terms of the conventional single self-energy
does not hold in the MI. So what can be deduced about $I_{L}$  in this case?

 This is easily answered by repeating the analysis of the charge $n$ as in sec.\ \ref{subsection:sec4A}, using
eq.\ \ref{eq:6} for $G(\w)$ expressed in terms of the usual single self-energy $\Sigma(\w)$. The resultant  
equation for $n$ (analogous to that just after eq.\ \ref{eq:22}) is then
\begin{equation}
\begin{split}
\tfrac{1}{2}n ~=&~ \tfrac{(-1)}{\pi}\mathrm{Im} \int_{-\infty}^{0}d\w ~
G(\w) \frac{\partial\Sigma(\w)}{\partial\w}
\\
&~+~ \tfrac{(-1)}{\pi}\mathrm{Im} \int_{-\infty}^{0}d\w ~
\frac{\partial}{\partial\w} \ln\left[\w^{+}-\epsilon_{d}^{\pd} - \Sigma(\w) - \tfrac{1}{4}t_{*}^{2}G(\w)\right] 
\\
&~+~\tfrac{1}{4}t_{*}^{2}~ \tfrac{(-1)}{\pi}\mathrm{Im} \int_{-\infty}^{0}d\w ~
G(\w)\frac{\partial G(\w)}{\partial\w}.
\end{split}
\nonumber
\end{equation}
The first term here is simply $-\tfrac{1}{\pi}I_{L}$ (eq.\ \ref{eq:11}), the final term is again given by 
$\tfrac{1}{4}t_{*}^{2}D(0)G^{R}(0)$ and so vanishes in the MI; and the middle term follows simply, to give 
\begin{subequations}
\label{eq:52}
\begin{align}
\tfrac{1}{2}n ~=&~-\tfrac{1}{\pi}I_{L} +\left[
1~-~ \tfrac{1}{\pi}\mathrm{tan}^{-1}\left(
\frac{0^{+}}{-\tilde{\epsilon}_{d}^{*}}
\right)
\right]
\\
=&~-\tfrac{1}{\pi}I_{L} ~+~\theta\left(
-\tilde{\epsilon}_{d}^{*}
\right).
\end{align}
\end{subequations}
Here, $\tilde{\epsilon}_{d}^{*}$ is a renormalised level defined naturally in terms of the standard \emph{single} self-energy by
(\emph{cf} eqs.\ \ref{eq:24},\ref{eq:25})
\begin{equation}
\label{eq:53}
\tilde{\epsilon}_{d}^{*}~=~ \epsilon_{d}^{\pd}~+~\Sigma^{R}(0)~+~\tfrac{1}{4}t_{*}^{2}G^{R}(0)
~=~\epsilon_{d}^{*}~+~\tfrac{1}{4}t_{*}^{2}G^{R}(0),
\end{equation}
with $\tilde{\epsilon}_{d}^{*} \neq 0$ throughout the generic ph-asymmetric ($\eta \neq 0$) Mott insulator 
(as shown below). But since $n =1$ throughout the MI, eq.\ \ref{eq:52} gives
\begin{equation}
\label{eq:54}
I_{L} ~= ~\tfrac{\pi}{2}~\left[\theta (-\tilde{\epsilon}_{d}^{*})~-~\theta (\tilde{\epsilon}_{d}^{*})\right].
\end{equation}
Throughout the generic ph-asymmetric MI, the Luttinger integral thus has constant magnitude 
$|I_{L}| = \frac{\pi}{2}$ \emph{independent of interaction 
strength}. The sign change in eq.\ \ref{eq:54} for $I_{L}$ also expected, since under a ph-transformation
it is easily shown that $\tilde{\epsilon}_{d}^{*}(\eta) = -\tilde{\epsilon}_{d}^{*}(-\eta)$
and $I_{L}(\eta) =-I_{L}(-\eta)$. 

The result $|I_{L}|=\tfrac{\pi}{2}$ generalises to the non-Fermi liquid Mott insulator the familiar Luttinger theorem applicable to the Fermi liquid metal, $I_{L}=0$. Precisely the same result is also  found for the local moment phases of a wide range of quantum impurity models \emph{per se}, as elaborated in ref.\ \onlinecite{LTG2014}.
And it likewise arises~\cite{LTG2014,Rosch_Lutt} trivially in the atomic (or `Hubbard atom') limit, $t_{*}=0$;  suggesting that it reflects perturbative continuity to that limit, in the same way that $I_{L}=0$ for the metallic Fermi liquid reflects 
adiabatic continuity to the non--interacting limit $U=0$.

Since eq.\ \ref{eq:19} relates the single self-energy to the two-self-energies, 
the renormalised level $\tilde{\epsilon}_{d}^{*}$ (eq.\ \ref{eq:53})  is of course readily related to the
$\tilde{\epsilon}_{d\sigma}^{*}$ (eq.\ \ref{eq:25}) defined in terms of the two-self-energies, on which we have focused in preceding sections. From eqs.\ \ref{eq:19},\ref{eq:25},\ref{eq:53} it follows
directly that
\begin{equation}
\label{eq:55}
\tilde{\epsilon}_{d}^{*}~=~ \tfrac{1}{2}\left[\tilde{\epsilon}_{d\uparrow}^{*}+\tilde{\epsilon}_{d\downarrow}^{*}\right] -
\tfrac{1}{2}\left[\tilde{\epsilon}_{d\uparrow}^{*}-\tilde{\epsilon}_{d\downarrow}^{*}\right]~
{\cal{P}}\Big(
\frac{1}{\tilde{\epsilon}_{d\uparrow}^{*}+\tilde{\epsilon}_{d\downarrow}^{*}}
\Big)
\end{equation}
(where ${\cal{P}}$ denotes a principal value). At ph-symmetry $\eta=0$, where 
$\tilde{\epsilon}_{d\uparrow}^{*}+\tilde{\epsilon}_{d\downarrow}^{*}=0$, it follows that
$\tilde{\epsilon}_{d}^{*}=0$ throughout the MI. Hence from eq.\ \ref{eq:54} (or eq.\ \ref{eq:52}a), $I_{L}=0$ 
precisely at ph-symmetry, so that in this limit $I_{L}=0$ throughout \emph{both} the metallic and MI phases
(as arises also in the atomic limit~\cite{LTG2014,Rosch_Lutt}).

The ph-symmetric limit aside, however, eq.\ \ref{eq:55} yields
$\tilde{\epsilon}_{d}^{*} =2\tilde{\epsilon}_{d\uparrow}^{*}\tilde{\epsilon}_{d\downarrow}^{*}
/(\tilde{\epsilon}_{d\uparrow}^{*}+\tilde{\epsilon}_{d\downarrow}^{*})$, i.e.\
\begin{equation}
\label{eq:56}
\frac{1}{\tilde{\epsilon}_{d}^{*}}~=~ \frac{1}{2}\left(
\frac{1}{\tilde{\epsilon}_{d\uparrow}^{*}}~+~\frac{1}{\tilde{\epsilon}_{d\downarrow}^{*}}
\right).
\end{equation}
Hence, since $\tilde{\epsilon}_{d\downarrow}^{*}>0$ and $\tilde{\epsilon}_{d\uparrow}^{*}<0$ throughout the MI
(eq.\ \ref{eq:29}), 
$\tilde{\epsilon}_{d}^{*} >0$ for
$|\tilde{\epsilon}_{d\downarrow}^{*}| <|\tilde{\epsilon}_{d\uparrow}^{*}|$, and \emph{vice versa} (with 
$\tilde{\epsilon}_{d}^{*}$ thus generically non-zero as asserted above).


\subsection{Renormalised level $\epsilon_{d}^{*}$}
\label{subsection:sec4F}

One can also determine the behaviour in the MI of the interaction-renormalised levels
$ \epsilon_{d}^{*}=\epsilon_{d}^{\pd}+\Sigma^{R}(0)$, given in terms of the single self-energy just as
in the metallic phase (eq.\ \ref{eq:9}); which again shows a  difference between ph-symmetry $\eta=0$ and 
the generic asymmetric case.

Using eq.\ \ref{eq:5}b, one can repeat the arguments of sec.\ \ref{section:sec3} to obtain an expression for the charge 
$n$, except that the conventional Luttinger integral $I_{L}$ is non-vanishing in the MI. The result is precisely 
eq.\ \ref{eq:10} again, 
\begin{subequations}
\label{eq:57}
\begin{align}
\tfrac{1}{2}n~=&~\int_{-\infty}^{-\epsilon_{d}^{*}} d\epsilon ~ \rho_{0}^{\pd}(\epsilon)~-~
\tfrac{1}{\pi}I_{L}
\\
=&~\int_{-\infty}^{-\epsilon_{d}^{*}} d\epsilon ~ \rho_{0}^{\pd}(\epsilon)
~+~\tfrac{1}{2}~\left[\theta (\tilde{\epsilon}_{d}^{*})~-~\theta (-\tilde{\epsilon}_{d}^{*})\right]
\end{align}
\end{subequations}
with eq.\ \ref{eq:54} for $I_{L}$ used in the second line.
The free lattice density of states $\rho_{0}(\epsilon)$ is given by eq.\ \ref{eq:2}, and has band edges at
$\epsilon = \pm t_{*}$. Since $n=1$ throughout the MI, it follows from eq.\ \ref{eq:57}b that\\
\noindent (a) for $\tilde{\epsilon}_{d}^{*}>0$, the renormalised level $\epsilon_{d}^{*} >+t_{*}$, while
for $\tilde{\epsilon}_{d}^{*}<0$, the level satisfies $\epsilon_{d}^{*} <-t_{*}$;\\
\noindent (b) for $\tilde{\epsilon}_{d}^{*}=0$ by contrast -- as occurs at ph-symmetry -- $\epsilon_{d}^{*} =0$.

Precisely at ph-symmetry, the renormalised level $\epsilon_{d}^{*}$ thus
vanishes throughout the MI phase, as well as throughout the metallic phase (sec.\ \ref{section:sec3}); and as such 
exhibits no signature of the Mott transition.
Away from ph-symmetry by contrast, however close, $|\epsilon_{d}^{*}| >t_{*}$ throughout the MI; while in the metallic phase 
$\epsilon_{d}^{*}$ vanishes as the transition is approached (sec.\ \ref{section:sec3}). In the generic case, therefore, the Mott transition is evident in the discontinuity in $|\epsilon_{d}^{*}|$ as the transition is approached; and which behaviour is indeed seen in the NRG results of fig.\ \ref{fig:fig4}. 


\section{NRG results}
\label{section:sec5}

In addition to the analytic results of previous sections, we provide further numerical results obtained 
by solving the DMFT self-consistency equation (eq.\ \ref{eq:6}) using the full density matrix 
generalisation~\cite{PetersPruschkeAnders2006,*WeichselbaumDelft2007} of NRG.~\cite{NRGrmp}

\begin{figure}
\includegraphics{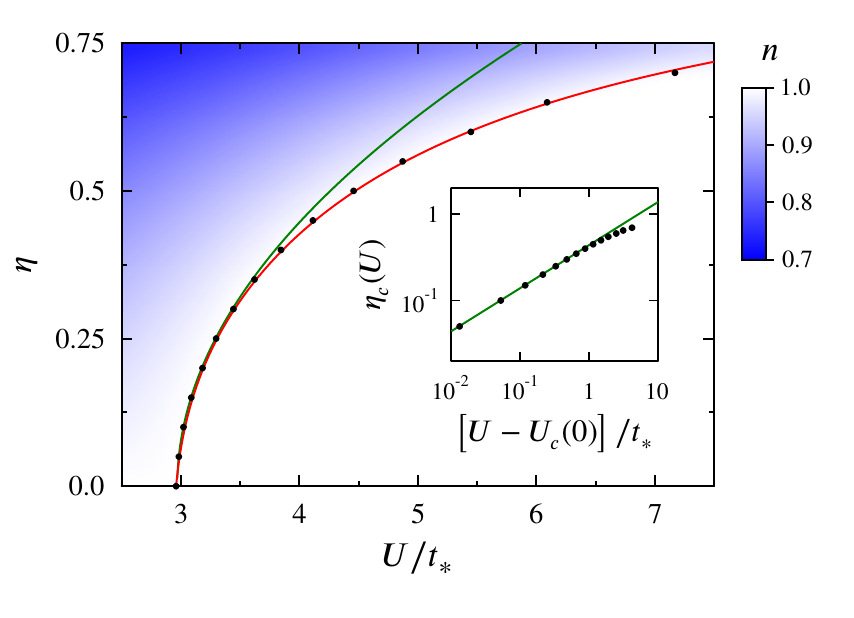}
\caption{\label{fig:fig5} 
NRG determined phase boundary (points) in the $(U,\eta)$-plane, showing the critical line $\eta_{c}(U)$
(or equivalently  $U_{c}(\eta)$). Charge $n$ in the metal is colour-coded as indicated;
$n=1$ throughout the MI. Green line shows the behaviour $\eta_{c}(U) \sim (U-U_{c}(0))^{1/2}$ arising for 
$U \rightarrow U_{c}(0)$. The latter is seen most clearly in the inset, showing data on a log-log plot, with 
solid line the square root form. Red line shows the empirical function given from solution of eq.\ \ref{eq:59}. 
}
\end{figure}

Fig.\ \ref{fig:fig5} shows the NRG phase boundary for the critical line $\eta_{c}(U)$, or equivalently
$U_{c}(\eta)$ ($=U_{c}(-\eta)$); determined by locating the points in the  $(U,\eta)$-plane  where $n\rightarrow 1$ on 
approaching the transition from the metal (recall that $n<1$ throughout the metal for all $\eta > 0$).
$n$ itself is calculated using eq.\ \ref{eq:4}. For the ph-symmetric point $\eta =0$, we find 
$U_{c}(0)/t_{*} \simeq 2.95$, in very good agreement with previous work.~\cite{BullaPRL99}

The results of fig.\ \ref{fig:fig5} establish clearly (see inset) that as the ph-symmetric point is approached, 
$\eta_{c}(U)$ vanishes with an exponent of $\tfrac{1}{2}$,
\begin{equation}
\label{eq:58}
\eta_{c}(U) ~\overset{U \rightarrow U_{c}(0)}{\sim} ~ \left(U-U_{c}(0)\right)^{\zeta}~~~~:~\zeta = \tfrac{1}{2}
\end{equation}
(and which is seen to be well satisfied in practice for $\eta_{c}(U) \lesssim 0.25$ or so). This  
is also consistent with a previous estimate of the exponent using a diagrammatic Monte Carlo method.~\cite{WernerMillis2007}

Fig.\ \ref{fig:fig5} further shows comparison of the NRG results for $\eta_{c}(U)$ to the following equation
\begin{equation}
\label{eq:59}
\tilde{U} - \tilde{U}_{c}(0) = e^{\phi}\Big[
\frac{1}{1-\eta_{c}}e^{-\phi (1-\eta_{c})\tilde{U}} ~-~e^{-\phi\tilde{U}_{c}(0)}
\Big],
\end{equation}
where $\tilde{U} = U/W$ (with $W=2t_{*}$ the full width of the free density of states 
$\rho_{0}(\epsilon)$, eq.\ \ref{eq:2}), and $\phi$ is a single adjustable constant (with $\phi =8$ taken in 
fig.\ \ref{fig:fig5}). Solution of eq.\ \ref{eq:59} gives $\eta_{c}$ as a function of $\tilde{U}$ for  
$\tilde{U}\geq \tilde{U}_{c}(0)$; for $\tilde{U} \rightarrow \infty$ in particular it is readily seen to yield
$\eta_{c}(U) \sim 1-1/\tilde{U} = 1-W/U$ (in agreement with the argument of 
sec.\ \ref{subsection:PD} that its asymptotic behaviour must be  $\eta_{c}(U) \sim 1-bW/U$ with $b \geq 1$, 
although our NRG results are insufficient to conclude whether $b=1$ or $>1$).
Eq.\ \ref{eq:59} is simply empirical. As seen from fig.\ \ref{fig:fig5} however, it gives a rather good description 
of the data over essentially the full range; albeit that for sufficiently small $\eta_{c}$ close to ph-symmetry, 
it is in fact ultimately linear (i.e. gives eq.\ \ref{eq:58} with $\zeta =1$ instead of $\tfrac{1}{2}$).

As the phase boundary is approached from the metallic phase, the charge $n\rightarrow 1$ from below. On approaching 
it by increasing $U$ towards $U_{c}(\eta)$ at fixed asymmetry $\eta$, we indeed find the expected asymptotic
behaviour eq.\ \ref{eq:42}; 
\emph{viz} $1-n \sim g~ (U_{c}-U)$ with exponent unity, and with a constant $g$ which vanishes as 
$\eta \rightarrow 0$ (obviously so, since $n=1$ for all $U$ at ph-symmetry $\eta =0$).

Likewise, on approaching the phase boundary by decreasing $\eta$ towards $\eta_{c}(U)$ for fixed interaction 
$U \geq U_{c}(0)$, we find the leading asymptotic behaviour $1-n \sim a~(\eta-\eta_{c})$ with an exponent of unity. The coefficient $a$ here is of course $a=-(\partial n/\partial \eta)_{\eta =\eta_{c}+}$, and hence (from eq.\ \ref{eq:3} for 
$\eta$), $a= \tfrac{U}{2} \chi_{c}(\eta =\eta_{c}+)$ where $\chi_{c} = -\partial n/\partial \epsilon_{d}  \geq 0$ is the 
charge susceptibility.~\cite{fnchargesusc} In otherwords,
\begin{equation}
\label{eq:60}
1-n~\overset{\eta \rightarrow \eta_{c}(U)+}{\sim}~ \tfrac{U}{2}\chi_{c}^{\pd}(\eta =\eta_{c}+)~~
(\eta-\eta_{c})
\end{equation}
with $\chi_{c}(\eta =\eta_{c}+)$ the charge susceptibility of the metal at the transition. 
Since this coefficient is finite for generic $U >U_{c}(0)$ (i.e.\ generic ph-asymmetry $\eta >0$) -- and since $\chi_{c}=0$ throughout the incompressible Mott insulator -- the charge susceptibility is thus in general discontinuous across the Mott transition. The sole exception is for  $U=U_{c}(0)$, where the transition occurs at the ph-symmetric point $\eta_{c} =0$;
for which we find the coefficient $a =\tfrac{U}{2} \chi_{c}(\eta =\eta_{c}+) =0$, i.e.\ 
a vanishing susceptibility at the transition. These results are in agreement with those from a diagrammatic Monte Carlo 
study.~\cite{WernerMillis2007} As shown further in Appendix \ref{section:scalingphi}, our data are also consistent with the scaling form~\cite{WernerMillis2007,fnscalingform}
\begin{equation}
\label{eq:61}
1-n~\overset{\eta \rightarrow \eta_{c}(U)+}{\sim}~ (\tilde{U}\eta)^{2}~\Phi\Big( \frac{\eta}{\eta_{c}(U)}\Big)
\end{equation}
where $\Phi(y)$ vanishes for $y=1$, but has a non-zero first derivative,
and tends to a constant as $y\rightarrow \infty$. Eq.\ \ref{eq:61} generates precisely eq.\ \ref{eq:60}, with a
$\chi_{c}(\eta =\eta_{c}+) \propto \eta_{c}(U)$ which is thus in general non-zero at the transition, vanishing only at the ph-symmetric point.


\subsection{Single-particle dynamics and scaling}
\label{subsection:sec5A}

\begin{figure}
\includegraphics{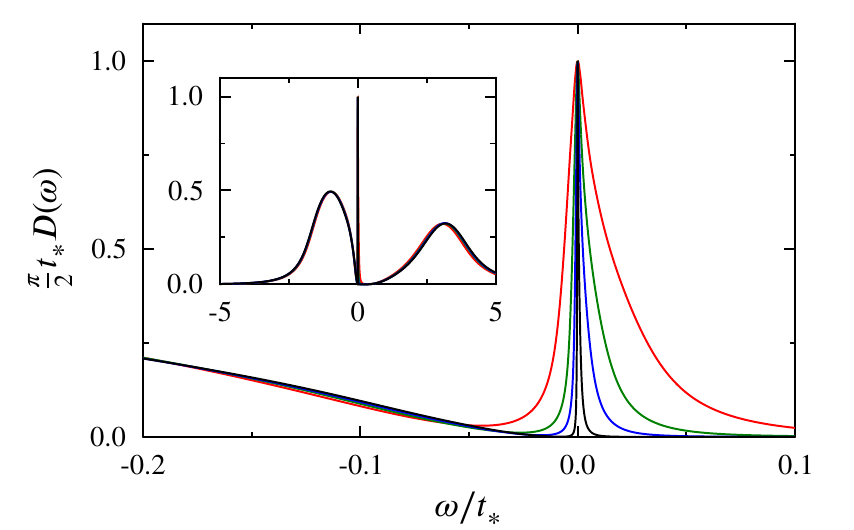}
\caption{\label{fig:fig6} 
For fixed $\eta =0.5$, single-particle spectrum $\tfrac{\pi}{2}t_{*}D(\w)$ \emph{vs} $\w/t_{*}$ 
on approaching the transition  ($U_{c}(\eta)\simeq 4.48t_{*}$) from the metal, for $U/t_{*} = 4.3$ (red line), 
$4.4$ (green), $4.44$ (blue) and $4.47$ (black). Inset shows $D(\w)$ on all scales, including Hubbard bands.
Main panel shows a close-up of the low-energy Kondo resonance, which progressively narrows and vanishes `on the spot' as
$U \rightarrow U_{c}(\eta)-$. 
}
\end{figure}

A representative example of single-particle dynamics is given in fig.\ \ref{fig:fig6}. For fixed asymmetry 
$\eta = 1+2\epsilon_{d}/U =0.5$, the spectrum $D(\w)$  is shown for the metallic phase, on increasing $U$ progressively towards the transition ($U_{c}(\eta)/t_{*} \simeq 4.48$). The general behaviour is as outlined in sec.\ \ref{subsection:PD}. The spectrum on `all scales' (fig.\ \ref{fig:fig6} inset) consists of upper and lower Hubbard bands, and a low-energy Kondo resonance straddling the Fermi level $\w =0$ (which lies fairly close to the upper edge of the lower Hubbard band for the $\eta$ shown). As the transition is approached the Kondo resonance narrows progressively (fig.\ \ref{fig:fig6} main panel), and
vanishes `on the spot' as $U \rightarrow U_{c}(\eta)-$.

The width of the resonance is of course characterised by a low-energy Kondo scale $\w_{\mathrm{K}}$ 
(proportional to the quasiparticle weight $Z = [1-(\partial\Sigma^{R}(\w)/\partial\w)_{\w =0}]^{-1}$); which we define
in practice as the right half-width (i.e.\ $\w >0$) at half-maximum of $D(\w)$. As the transition is approached from the metallic phase, $\w_{\mathrm{K}} \equiv \w_{\mathrm{K}}(U,\eta)$ vanishes, and we find it does so with exponent unity 
regardless of the direction of approach; i.e.\
\begin{equation}
\label{eq:62}
(Z \propto) ~~~\w_{\mathrm{K}}~~{\sim}~~ c \left(\eta-\eta_{c}\right)+c^{\prime}\left(U_{c}-U\right)
\end{equation}
close enough to any point $(U_{c},\eta_{c})$ on the phase boundary (including the ph-symmetric point where $\eta_{c}=0$).

\begin{figure}
\includegraphics{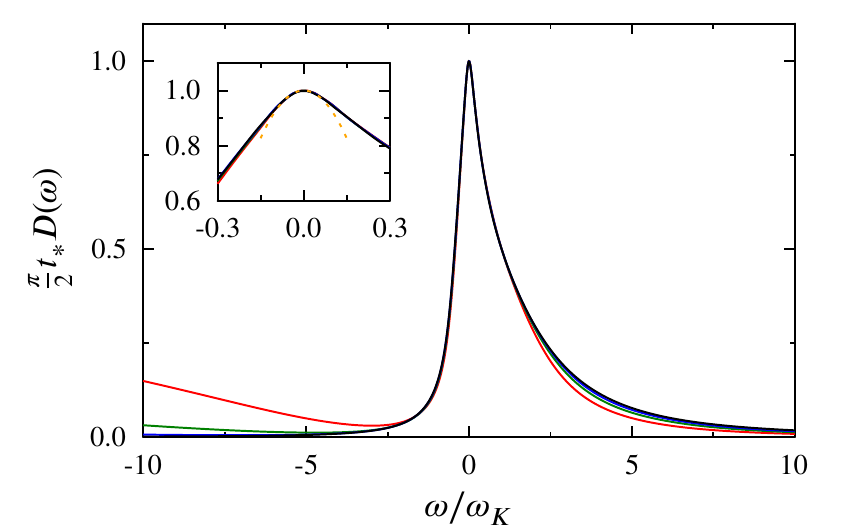}
\caption{\label{fig:fig7} 
Universl spectral scaling on approaching the transition from the metal, for fixed $\eta $ ($=0.5$) on increaing $U$
towards $U_{c}(\eta)$. $\tfrac{\pi}{2}t_{*}D(\w)$ is shown \emph{vs} $\w/\w_{\mathrm{K}}$,
with $\w_{\mathrm{K}}$ the low-energy Kondo scale, for for the same $U/t_{*}$ as fig.\ \ref{fig:fig6}.
Clear universality arises. The Fermi level spectrum $\tfrac{\pi}{2}t_{*}D(0) =1$ as required from eq.\ \ref{eq:13}.
\emph{Inset}: close-up of top of scaling spectrum showing quadratic behaviour in $\w/\w_{\mathrm{K}}$ (dashed line), further indicating emergent ph-symmetry as the transition is approached.
}
\end{figure}

Since $\w_{\mathrm{K}}$ vanishes on approaching the transition, one expects universal scaling of the spectrum 
in terms of $\w/\w_{\mathrm{K}}$; although the existence of a vanishing low-energy scale is \emph{not} by itself sufficient to guarantee spectral scaling as an entire function of $\w/\w_{\mathrm{K}}$.
We find that such universal scaling takes place only when the transition is approached by varying $U$ for \emph{fixed} asymmetry $\eta$. That this indeed arises is seen clearly in fig.\ \ref{fig:fig7} (for fixed $\eta =0.5$), where 
$\tfrac{\pi}{2}t_{*}D(\w)$ ($=D(\w)/\rho_{0}(0)$) is shown versus $\w/\w_{\mathrm{K}}$ for the same interaction strengths used in fig.\ \ref{fig:fig6}. Clear spectral scaling is seen to arise, and occurs to increasingly larger values of $|\w|/\w_{\mathrm{K}}$ ($\gg 1$) as the transition is progressively approached and $\w_{\mathrm{K}}$ vanishes.
We find this behaviour to be generic: the universal scaling spectra 
$\tfrac{\pi}{2}t_{*}D(\w)  \equiv {\cal{D}}(\w/\w_{\mathrm{K}}; \eta)$
form a family of $\eta$-dependent functions (with that at ph-symmetry naturally symmetric 
about $\w/\w_{\mathrm{K}}=0$). Note also from fig.\ \ref{fig:fig7} that the spectrum at the Fermi level 
$\w =0$ clearly approaches $\tfrac{\pi}{2}t_{*}D(\w=0) =1$, i.e.\ $D(\w=0) =\rho_{0}(0)$; as indeed required by 
eq.\ \ref{eq:13}, and indicative of the emergent ph-symmetry that arises on approaching the transition for the generic ph-asymmetric model (sec.\ \ref{section:sec3}). This is further evident from the inset to 
fig.\ \ref{fig:fig7}, where the scaling spectrum at low-energies is seen to have the purely quadratic behaviour 
$D(\w)-D(0) \propto -(\w/\w_{\mathrm{K}})^{2}$ that is characteristic of ph-symmetry.

If by contrast the transition is approached at fixed $U$ by decreasing $\eta$, then universal spectral scaling as a  
function of $\w/\w_{\mathrm{K}}$ (or equivalently $\w/Z$) does not occur, except on the lowest scales 
$|\w|/\w_{\mathrm{K}} \ll 1$ where it is of course guaranteed by Fermi liquid theory. This is why full universal scaling in terms of $\w/Z$ was not seen in ref.\ \onlinecite{Zitko+BSS2013}, where the transition was approached in this way. \\

Finally, in regard to approaching the transition by decreasing $\eta$ at fixed $U>U_{c}(0)$, one further point bears note.
While it is physically natural to consider the dependence of physical properties on the doping $\delta :=1-n$ (which may be controlled in experiment), this dependence can become rather complicated for low-doping near $U=U_{c}(0)$, due to the vanishing of the charge susceptibility at the ph-symmetric point. We illustrate the point with reference to the quasiparticle weight $Z$, in order to understand the results of ref.\ \onlinecite{Zitko+BSS2013} for the doping dependence of $Z/\delta$. Sufficiently close to the transition, 
\begin{equation}
\label{eq:63}
\delta ~=~1-n~\sim ~ a\left(\eta-\eta_{c}\right)~+~b\left(\eta-\eta_{c}\right)^{2}
\end{equation}
where (as in eq.\ \ref{eq:60}) $a =\tfrac{U}{2} \chi_{c}(\eta =\eta_{c}+)$ with $\chi_{c}$ the charge susceptibility.
At $U=U_{c}(0)$ precisely, $a=0$ as discussed above, and thus $\delta \propto (\eta -\eta_{c})^{2}$ (where $\eta_{c}=0$ at this ph-symmetric point). But as in eq.\ \ref{eq:62}, the quasiparticle weight $Z \propto (\eta -\eta_{c})$, whence
$Z \propto \delta^{1/2}$. The quantity $Z/\delta$ thus diverges as $\delta \rightarrow 0$ for $U=U_{c}(0)$.

For any $U>U_{c}(0)$ by contrast, the charge susceptibility at the transition is finite (i.e.\ $a>0$), so 
$\delta \propto (\eta -\eta_{c})$. Since $Z$ vanishes linearly in $(\eta -\eta_{c})$, it likewise vanishes
linearly with doping $\delta$, as indeed observed for large enough $U$ in  ref.\ \onlinecite{Zitko+BSS2013} (fig.\ 1(c)).
Notice however from eq.\ \ref{eq:63} that this behaviour sets in only when $\delta \ll a^{2}/4b$. Close to $U=U_{c}(0)$ the charge susceptibility (and hence $a$) is very small, so the behaviour $Z \propto \delta$ may occur only over an extremely narrow range of doping $\delta$. This is indeed as found in fig.\ 1(c) of ref.\ \onlinecite{Zitko+BSS2013} for $U/t_{*}=3$ -- just slightly larger than $U_{c}(0)$ -- where this linear behaviour is not seen at all: instead, $Z/\delta$ appears to diverge, because of the dominance of the quadratic term in eq.\ \ref{eq:63}.


\comment{

\section{Concluding remarks}
\label{section:sec6}

}

\appendix  


\section{Scaling function $\Phi$}
\label{section:scalingphi}

\begin{figure}
\includegraphics{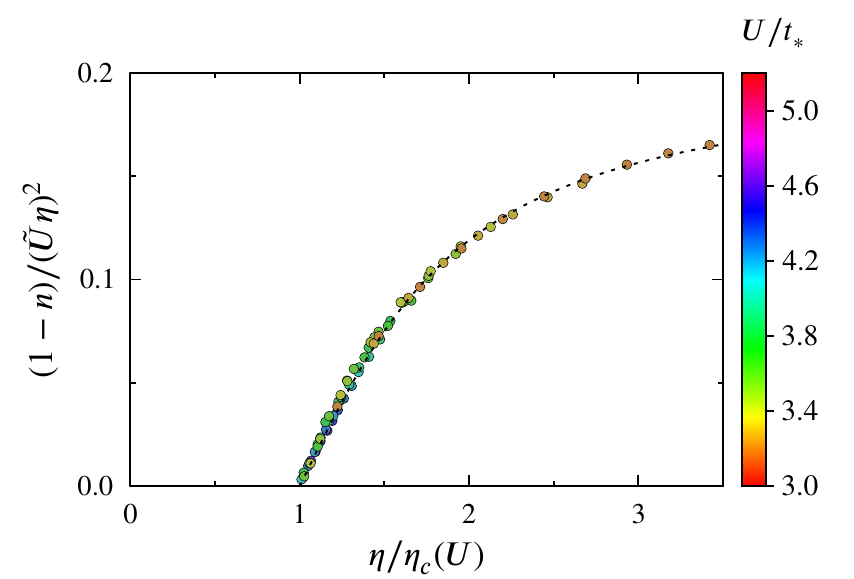}
\caption{\label{fig:figA1} 
$(1-n)/(\tilde{U}\eta)^2$ \emph{vs} $\eta/\eta_c(U)$ for different values of $U/t_{*}$ (indicated by the colour scale), 
yielding the scaling function $\Phi(\eta/\eta_c(U))$. Dashed line shows the empirical form eq.\ \ref{eq:A1}.
}
\end{figure}

Our NRG results for $1-n$ on approaching the transition from the metal exhibit the scaling behaviour 
eq.\ \ref{eq:61},~\cite{WernerMillis2007,fnscalingform} as stated in sec.\ \ref{section:sec5}.
We demonstrate this in the following way. For a given $U>U_c(0)$ the charge is calculated as a function of $\eta$, close to the phase boundary $\eta_c(U)$. From this, one obtains  the quantity $(1-n)/(\tilde{U}\eta)^2$ as a function of $\eta/\eta_c(U)$. According to eq.\ \ref{eq:61}, these data should scale onto a universal curve $\Phi(\eta/\eta_c(U))$, independently of $U$. This is indeed the case, as shown in fig.\ \ref{fig:figA1}. 
The data collapse occurs over a wider range of $\eta/\eta_c(U)$ as $U$ decreases toward $U_c(0)$, where $\eta_c(U)$ vanishes.
The function $\Phi(y)$ is seen to have the properties discussed in the text after eq.\ \ref{eq:61}, and as seen in
fig.\ \ref{fig:figA1} is in fact well described by the empirical form
\begin{equation}
\label{eq:A1}
\Phi(y) ~=~ p~ \mathrm{tan}^{-1}[q(y-1)],
\end{equation}
with constants $p=0.13$ and $q =1.3$.

\begin{acknowledgments}
We are grateful to the EPSRC for financial support, under grant EP/I032487/1.
\end{acknowledgments}

\bibliography{arxiv}

\begin{thebibliography}{57}
\bibitem [{fnp()}]{fnphyssoln1}%
  \BibitemOpen
  \href@noop {} {}\bibinfo {note} {The formal alternative, $\epsilon_{d}^{*}
  +\tfrac{1}{4}t_{*}^{2}G(0)=0$, correponds to $G(0)$ pure real, and is thus
  inadmissible in the metal.}\BibitemShut {Stop}%
\bibitem [{fns()}]{fnscalingform}%
  \BibitemOpen
  \href@noop {} {}\bibinfo {note} {Note that with $\mu \equiv -\epsilon_{d}$
  the chemical potential, and $\mu_{1} := U/2$, the asymmetry $\eta$ (eq.\
  \ref{eq:3}) reads $\eta = -\tfrac{2}{U}(\mu-\mu_{1})$, and the scaling form
  eq.\ \ref{eq:61} may be written as $1-n \sim
  \tfrac{1}{t_{*}^{2}}(\mu-\mu_{1})^{2}\Phi(\tfrac{\mu-\mu_{1}}{\mu_{c}-\mu_{1}})$,
  as in ref.\ [\onlinecite{WernerMillis2007}].}\BibitemShut {Stop}%
\end{thebibliography}

\end{document}